\documentclass[twocolumn]{jpsj2}

\usepackage{color}

\newcommand{\bvec}[1]{\mib{#1}}

\makeatletter
\def\tr{\mathop{\operator@font tr}\nolimits}
\makeatother
\makeatletter
\def\Tr{\mathop{\operator@font Tr}\nolimits}
\makeatother

\title{
Slow relaxation in
the two dimensional electron plasma
under the strong magnetic field%
}

\def\runauthor{Ryo Kawahara and Hiizu Nakanishi}

\author{
 Ryo \textsc{Kawahara}\thanks{E-mail : ryokawa@stat.phys.kyushu-u.ac.jp} 
 and
 Hiizu \textsc{Nakanishi}\thanks{E-mail : naka4scp@mbox.nc.kyushu-u.ac.jp}}

\recdate{\today}

\kword{
non-neutral plasma,
two-dimensional turbulence,
numerical simulation,
long range force,
non-extensive system,
slow relaxation,
anomalous diffusion,
Levy flight
}

\inst{
Department of Physics, Kyushu University 33,
Fukuoka 812-8581.
}

\abst{%
We study slow relaxation processes in the point vortex model
for the two-dimensional
pure electron plasma
under the strong magnetic field. By numerical simulations,
it is shown that, 
from an initial state, the system undergoes the fast relaxation
to a quasi-stationary state, and then
goes through the slow relaxation to reach a final
state.
From analysis of simulation data,
we find
(i)
 the time scale of the slow relaxation
increases linearly to the number of electrons
if it is measured by the unit of
the bulk rotation time,
(ii)
 during the slow relaxation process,
 each electron undergoes an superdiffusive motion,
and
(iii)
the superdiffusive motion can be regarded as the
Levy flight, whose step size distribution is of the power law.
The time scale that each electron diffuses over the system
size turns out to be much shorter than that of the
slow relaxation, 
which suggests that the correlation among the
superdiffusive trajectories is important in the slow relaxation process.

}

\begin{document}

\maketitle

\section{Introduction}
\label{s_introduction}
The system of pure electron plasma under the strong magnetic field has
been of scientific interest for various aspects; 
Such a system does not satisfy the basic premise of the 
Boltzmann statistics, i.e. the existence of a subsystem weakly coupled
to the rest,
because the 
interaction between the constituent particles is long-range,
consequently
 any part of the system interacts strongly with the rest.
Under a certain
experimentally realizable condition,
the system behaves like a two-dimensional (2-d) system and
 is shown to be described as a 2-d point vortex system with
a unit vorticity, whose continuum description reduces to the 2-d Euler
equation of the incompressible inviscid fluid.  
\cite{btwodim_turbulence_physicist}

Historically, the point vortex system
 is the system for which Onsager
developed the idea of negative temperature in his
statistical theory and predicted the equilibrium states
of the large vortex cluster in  the
relaxation of 2-d high Reynolds number fluid.
\cite{
 bstatistical_hydrodynamics,
 bonsager_theory}
His theory has been advanced
\cite{
 bnegative_temperature_states,
 bstatistical_mechanics_negative,
 bnonaxisymmetric_thermal_equilibria
},
 and has been examined by 
simulations of several hundred particles (or vortices)
to show that the
equilibrium state is the vortex cluster that maximizes the
one-body entropy.
\cite{
 bstatistics_line_vortices
}

%



In contrast with  these results,
in recent 2-d electron plasma experiments,
the system has been demonstrated to relax into
several kinds of ``stationary'' or ``metaequilibrium''
 states,
including the minimum enstrophy state
\cite{btwo_dim_turbulence_exp}
 and
the vortex crystal state.
\cite{btwo_dim_vortex_crystal}
These are quite different from the above equilibrium state.
There
have been applied several statistical theories to understand these
relaxed states, which includes
 the minimum enstrophy theory, 
\cite{btwo_dim_turbulence_exp}
 the fluid entropy theory,
\cite{
 brelaxation_towards_statistical_equilibrium,
 bfinal_equilibrium_state,
 bstatmech_eulereq_jupiter
}
 the Tsallis statistics,
\cite{btwo_dim_turbulence_tsallis}
 the regional entropy theory for the vortex crystal,
\cite{bregional_maximum_entropy}
 etc., but
none of these
theories have successfully described these states on a
general ground.
\cite{
 bquasi_stationary_states,
 bmax_entropy_vs_min_enstrophy,
 bmixing_thermal_equilibrium,
 btwodim_euler_unique_final
}

%

These two contradicting results suggest that 
the system is trapped in
some states before
reaching the equilibrium in
the experiments.
In fact, we show  in the present work
that these stationary states
are only quasi-stationary and relaxes into
the equilibrium state very slowly.
This process of slow relaxation
is the major subject of the present paper.

Actually, for the stellar system with gravitation,
which is another example with long-range interaction,
it has been known that there are 
two types of relaxation:
the violent relaxation and the collisional relaxation.
The violent relaxation
\cite{
 bstatistics_violent_relaxation,
 bstatistical_mechanics_gravitating}
 is the fast process
which is 
caused by the complex mixing due to
the macroscopic collective interaction.
The dynamics is 
described by the Vlasov equation and
the two-body ``collisions'' do not play any role.
In contrast, the collisional relaxation
\cite{
 bstatistical_mechanics_gravitating,
 bbrownian_motion_dynamical_friction}
 is the slow process
 caused by
the collisions of the individual particles
and the relaxation time scale is much longer than that of the
violent relaxation. 
Such a relaxation has already been found and analyzed in the
molecular dynamics simulations of self-gravitating system.
\cite{
 blongterm_evolution_stellar_gravitating,
 bself-gravitating_stellar_systems
}
The separation of these two
relaxations is considered to be peculiar to the systems
with long-range interaction.

In the present system of the 2-d electron plasma
under the strong magnetic field,
there are also a fast and a slow relaxation.
The fast 
relaxation from an initial state takes place
through the violent motion of the stretching 
and folding in the density field
due to the macroscopic collective motion,
 and leads the system
to a quasi-stationary state. 
The dynamics is described by the 2-d Euler equation
and the individual two-body collisions do not play any role.
The typical time scale
for this fast relaxation is the rotation time of the bulk 
of the system $\tau_{\mathrm{rot}}$.


In contrast,
the relaxation
after the quasi-stationary states towards the equilibrium
is much slower than the fast relaxation.

Chavanis has presented a theoretical analysis on the dynamics of this
system
\cite{
 bkinetic_theory_of_point_vortices,
 bstatmech_twodim_vorices_stellar}
using the analogous idea developed in the self-gravitating system.
  He predicted that the ratio of the slow relaxation time to
the fast relaxation scales with the total number of particles $N$ as
$N / \ln N$, based on the idea that the relaxation takes place through the
diffusion of the individual particles caused by collisions,
which can be ignored in the violent relaxation.


For the \textit{neutral}
 plasma under the strong magnetic field, 
the theoretical analysis
by Taylor and McNamara
\cite{
 bplasma_diffusion_two_dimensions
}
has shown
that the individual particle motion converges to the normal
diffusive behavior only slowly,
and the limiting value of the diffusion coefficient have been found to
depend on the system size, which implies anomalous diffusion is
taking place.
Dawson et al. have pointed out that such behavior
of diffusion is attributed to
occasional long jumps of the particles convected by large vortices.
\cite{
 bnumerical_simulation_plasma_difusion_across
}


In some experiments of the 2-d fluid with vortices,
the diffusion of tracer particles have been often found to
be anomalous,
\cite{
 banomalous_diffusion_linear_array_vortices,
 banomalous_diffusion_trace_convection_rolls
}
and 
a Levy flight analysis has been proposed
\cite{bobservations_anomalous_diffusion}.

In this paper, 
we present the results of the numerical simulations on
the point vortex system.
We observe the $N$-dependence of the fast and slow
relaxation times and find that they are consistent with
those estimated by Chavanis. Special attention is on
the diffusion process during the slow relaxation,
for which we find an anomalous diffusion and analyze
it in terms of the Levy flight.


The paper is organized as follows:
In \S\ref{s_model_system_and_its_behavior}, we introduce
the 2-d electron plasma system under a magnetic field and define
the model.
In \S\ref{s_theories}, 
we briefly review some statistical theories.
The simulation method is explained in \S\ref{ss_method}
and results are described 
in comparison with the statistical theories in \S\ref{s_simulation}.
We analyze these results in terms of the Levy flight
in \S\ref{s_simulations_ss_analyses},
and the summary and
 discussion are given in \S\ref{s_conclusions_and_discussion}.

\section{Model system and its behavior}
\label{s_model_system_and_its_behavior}

\subsection{Equations of motion}
\label{ss_equations_of_motion}

\begin{figure}[b]
 \begin{center}
   \includegraphics[trim=20 0 0 0,width=0.40\columnwidth]{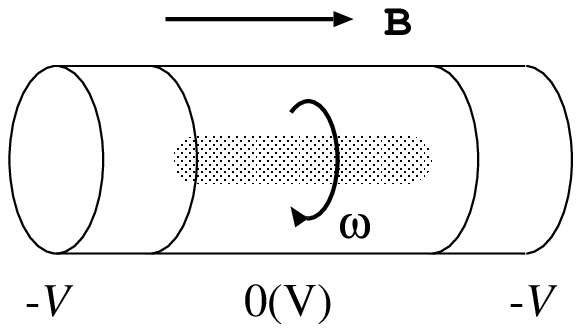}
  \caption{
   Schematic diagram of Malmberg trap of the pure electron plasma.}
  \label{f_malmberg}
 \end{center}
\end{figure}



The physical system we consider is the pure electron plasma in
a cylindrical container with the
strong magnetic field $\bvec{B}$ applied along the axis
(Fig. \ref{f_malmberg}).
In a certain situation,
\cite{btwo_dim_turbulence_exp, btwo_dim_vortex_crystal}
 it has been shown
that the interaction between the
electrons can be approximated by the
 two-dimensional Coulomb force
and the drift velocity $\bvec{v}_{i}$ of the $i$-th electron
in the plane perpendicular to $\bvec{B}$ is given by
\begin{equation}
 \label{m_drift}
 \mib{v}_{i} = \frac{\mib{E}(\mib{r}_{i})\times\mib{B}}{B^2} ,
\end{equation}
disregarding the cyclotron motion.
Here $\bvec{E}(\bvec{r})$ is the electric field
 and $\bvec{r}_{i}$
is the position of $i$-th electron in the plane.
Thus the electrons do not repel each other and can be contained in the
cylinder (Malmberg Trap).

After some normalization, it can be shown 
\cite{bquasi_stationary_states}
that the equations of motion
for the electron position $\bvec{r}_{i} = (x_{i},y_{i})$
are given by
\begin{equation}
\label{m_hamiltonian_eq}
 \frac{dx_{i}}{dt} = \frac{\partial H}{\partial y_{i}},
\quad
 \frac{dy_{i}}{dt} = -\frac{\partial H}{\partial x_{i}},
\end{equation}
with the Hamiltonian $H$,
\begin{equation}
\label{m_hamiltonian}
\begin{split}
  H
 &= - \frac{1}{2}\sum_{i}^{N}\sum_{j \neq i}^{N}
      G(\mib{r}_{i},\mib{r}_{j})
    - \frac{1}{2}\sum_{i}^{N}
      G_{\mathrm{m}}(\mib{r}_{i},\mib{r}_{i}) \\
 &= - \frac{1}{2}\sum_{i}^{N}\phi_{i}(\mib{r}_{i}),
\end{split}
\end{equation} 
where $G(\mib{r}_{i},\mib{r}_{j})$ is the 2-d Green function
for the electric potential
that satisfies
 $\nabla^{2}G(\mib{r},\mib{r}') = \delta(\mib{r} - \mib{r}')$
 with an appropriate boundary condition,
$G_{\mathrm{m}}(\mib{r}, \mib{r}')$ is the electric potential
at $\mib{r}$ brought about by the mirror charge 
induced by the charge at $\mib{r}'$,
 and $\phi_{i}(\mib{r})$ is the electric
potential due to all the electrons except for the $i$th one.
In the present case, we consider the system which is in a cylindrical
 container with the radius $R_{w}$, then $\phi_{i}$ is given by
 \cite{bnpoint_vortex, bdynamics_two_sign_point_vortices}
\begin{equation}
\label{mcylinder_potential}
\begin{split}
  \phi_{i}(\mib{r}) \equiv
 &+\frac{1}{2\pi}\sum_{j\neq i}^{N}
  \ln |\mib{z} - \mib{z}_{j}| \\
 &-\frac{1}{2\pi}\sum_{j}^{N}
  \left[
   \ln |z - \frac{R_{w}^{2}}{z_{j}^{*}}| 
   +\ln \frac{|z_{j}|}{R_{w}}
  \right],
\end{split}
\end{equation}
where $z = x + i y$ and 
$z^{*}$ is the complex conjugate of $z$;
the second term corresponds to the potential by
the mirror charges.

\subsection{Euler equation}
\label{sss_equations_of_motion_eularian}

Note that the density field 
\begin{equation}
 \label{m_density_delta}
 n(\mib{r}) = \sum_{i}\delta(\mib{r} - \mib{r}_{i}(t))
\end{equation}
with $\mib{r}_{i}(t)$ being a solution of 
eq. (\ref{m_hamiltonian_eq}) gives a singular solution
for the set of equations,
%
\begin{equation}
 \label{m_euler}
 \frac{Dn}{Dt} \equiv
 \frac{\partial n}{\partial t} + \mib{v}\cdot \nabla n = 0,
\end{equation}
\begin{equation}
 \label{m_drift_normalized}
 \mib{v} = \hat{\mib{z}}\times\nabla\phi = \left(
  -\frac{\partial \phi}{\partial y}, \frac{\partial \phi}{\partial x}
 \right),
\end{equation}
\begin{equation}
 \label{m_poisson}
\nabla^{2}\phi = n ,
\end{equation}
where $\hat{z}$ denotes the unit vector perpendicular to
the plane and $\nabla$ is the 2-d nabla.  

It can be shown that the density field $n$ is equal to 
the vorticity $\omega(\mib{r}) \equiv (\nabla \times \mib{v})_{z}$
of the 2-d
velocity field $\mib{v}$ and the velocity field $\mib{v}$ is
solenoidal ($\nabla \cdot \mib{v} = 0$),
therefore, the set of equations (\ref{m_euler})$-$(\ref{m_poisson})
 is equivalent to that of
the Euler equation for the 2-d incompressible inviscid fluid
with a free-slip (no-stress) boundary condition, but the
vorticity takes only a positive value in the present system.

To see the basic motion and the time scale, we
consider the ``pancake'' state
with the density field,
\begin{equation}
n(r) = \left\{
\begin{array}{ll}
 n_{0} & (r \leq R) \\
 0     & (r > R)
\end{array}
\right.
,
\end{equation}
which is a steady solution of eq.  (\ref{m_euler})$-$(\ref{m_poisson}).
The system shows a rigid rotation with
the period
\begin{equation}
\label{m_bulkrotation}
 \tau_{\mathrm{rot}}
 \equiv
 \frac{r}{v_{\mathrm{rot}}(r)}
  = \frac{4\pi}{n_{0}}
.
\end{equation}
This period gives the approximate time scale for the
macroscopic dynamics of
the system of  point vortices
with the average density
 $n_{0} \approx \pi N / R^{2}$.


If the system contains macroscopic numbers of electrons
that follow eq. (\ref{m_hamiltonian_eq}),
and their distribution
can be represented by a smooth density field $n_{\mathrm{m}}(x,y)$
obtained through averaging 
over a finite size mesh, i.e. coarse graining,
then the density field should follow
the following partial differential equations:

\begin{equation}
 \label{m_euler_collision}
 \frac{Dn_{\mathrm{m}}}{Dt} \equiv
 \frac{\partial n_{\mathrm{m}}}{\partial t}
  + \mib{v}_{\mathrm{m}} \cdot \nabla n_{\mathrm{m}} = C(n_{\mathrm{m}}),
\end{equation}
\begin{equation}
 \label{m_drift_normalized_collision}
 \mib{v}_{\mathrm{m}} = \hat{\mib{z}}\times\nabla\phi_{\mathrm{m}} = \left(
  -\frac{\partial \phi_{\mathrm{m}}}{\partial y}, \frac{\partial \phi_{\mathrm{m}}}{\partial x}
 \right),
\end{equation}
\begin{equation}
 \label{m_poisson_collision}
\nabla^{2}\phi_{\mathrm{m}} = n_{\mathrm{m}} .
\end{equation}
These are almost the same with
 eqs.
 (\ref{m_euler}) $-$ (\ref{m_poisson}),
but there is the extra term  $C(n_{\mathrm{m}})$
in the right hand side of eq. (\ref{m_euler_collision}).
This comes from the averaging on $n$ and
represents microscopic processes, thus
it is called the collision term, since it is 
analogous to the corresponding term in the Boltzmann equation
for the kinetic theory of gases.
Note that, in experimental situations, we observe the particles density
in a finite resolution,
therefore the averaging operations are essential in macroscopic observations.

When the system contains macroscopic number of particles,
 the collision term is small,
but it eventually causes the slow relaxation that leads the system
to the equilibrium state,
 and
estimating the effects of the collision term
is still under intense discussion.


\subsection{Constants of the dynamics}
\label{ss_constants_of_the_dynamics}
The Hamiltonian $H$ is a constant of the dynamics, and
is expressed in terms of the field variables as
\begin{equation}\label{m_energy}
\begin{split}
  H 
    &= -\frac{1}{2}\int d^{2}\mib{r}\int d^{2}\mib{r}' \,
        n(\mib{r})n(\mib{r}')G(\mib{r}, \mib{r}') \\
    &= -\frac{1}{2}\int d^{2}\mib{r} \, n(\mib{r})\phi(\mib{r}) \\
    &= \int d^{2}\mib{r} \, \frac{1}{2}\mib{v}^{2}(\mib{r}),
\end{split}
\end{equation}
which corresponds to the total energy.

In the case of a system with circular symmetry,
 the total angular momentum $I$ around the
center of the system,
\begin{equation}\label{m_angular_momentum}
\begin{split}
  I  
  &\equiv \sum_{i}^{N}r_{i}^{2} \\
  &= \int d^{2}\mib{r} \, r^{2}n(\mib{r}) \\
  &= \int d^{2}\mib{r} \, (\mib{r} \times \mib{v}(\mib{r}))_{\mathrm{z}},
\end{split} 
\end{equation}
is also a constant of the dynamics.

Since the 2-d Euler equation (\ref{m_euler}) $-$ (\ref{m_poisson})
represents an area-preserving dynamics, it
has a set of conserved quantities, called Casimir
constants,
\begin{equation}\label{m_casimirs}
  Z \equiv \int d^{2}\mib{r} \, f(n(\mib{r})),
\end{equation}
where $f(n)$ is an arbitrary function of the density field $n$.
In the point vortex system, however,
the Casimirs $Z$ cannot be defined except for 
the linear combination of the total number
of particles $N$
\begin{equation}
 \label{m_numberofparticles}
 N \equiv \int d^{2}\mib{r} \, n(\mib{r})
\end{equation}
since the density field $n$ has singularity of the delta function as in
 eq. (\ref{m_density_delta}).

\subsection{Stability and evolution of states}
\label{ss_stability_and_evolution_of_states}

\begin{figure}[b]
 \begin{center}
   \raisebox{0.19\columnwidth}{(a)}
   \hspace{-1.7em}
   \includegraphics[trim=40 0 100 0,width=0.22\columnwidth]
                   {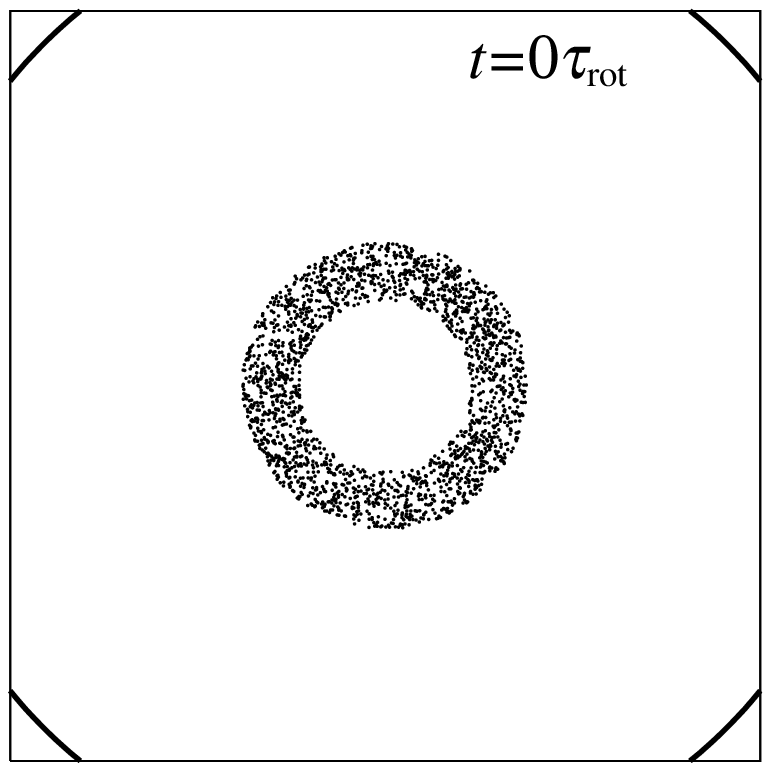}
   \includegraphics[trim=40 0 100 0,width=0.22\columnwidth]
                   {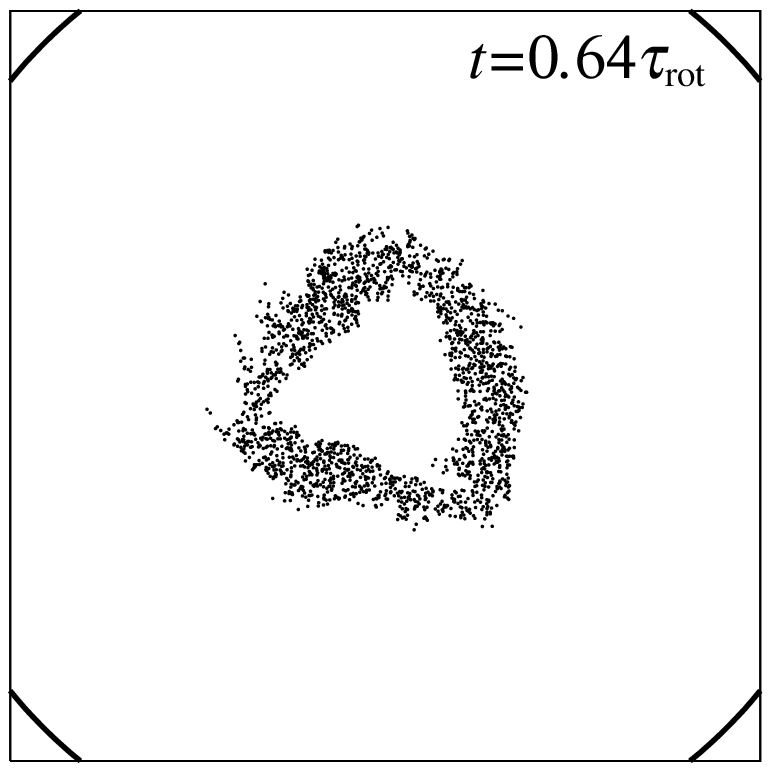}
   \includegraphics[trim=40 0 100 0,width=0.22\columnwidth]
                   {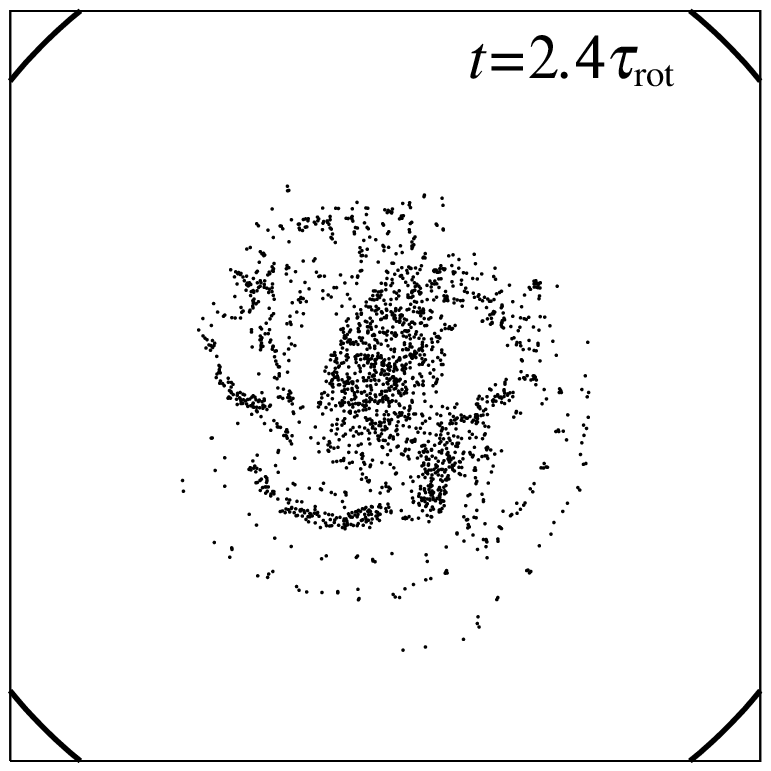}
   \includegraphics[trim=40 0 100 0,width=0.22\columnwidth]
                   {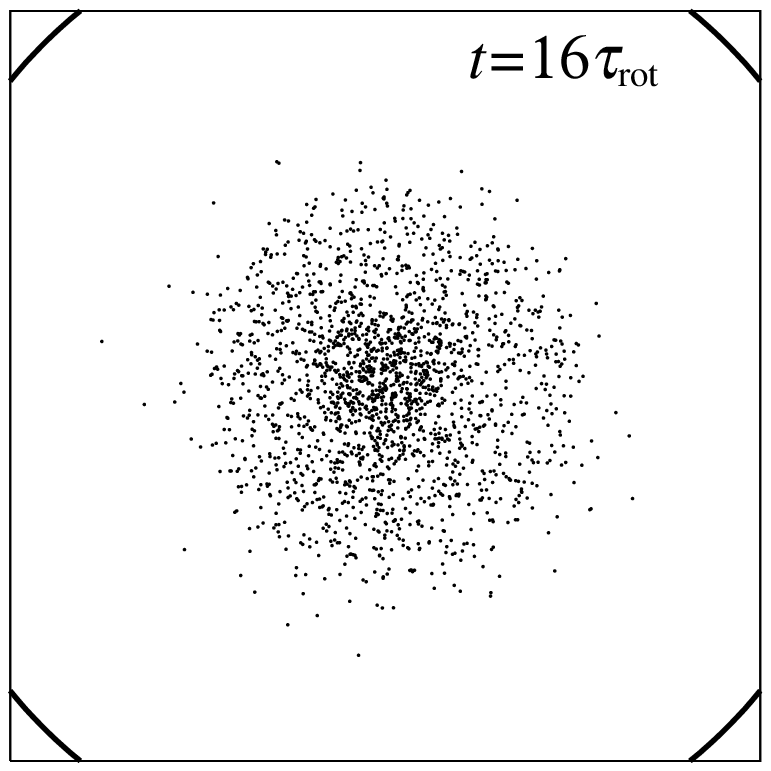}
   \\
   \raisebox{0.19\columnwidth}{(b)}
   \hspace{-1.7em}
   \includegraphics[trim=40 0 100 0,width=0.22\columnwidth]
                   {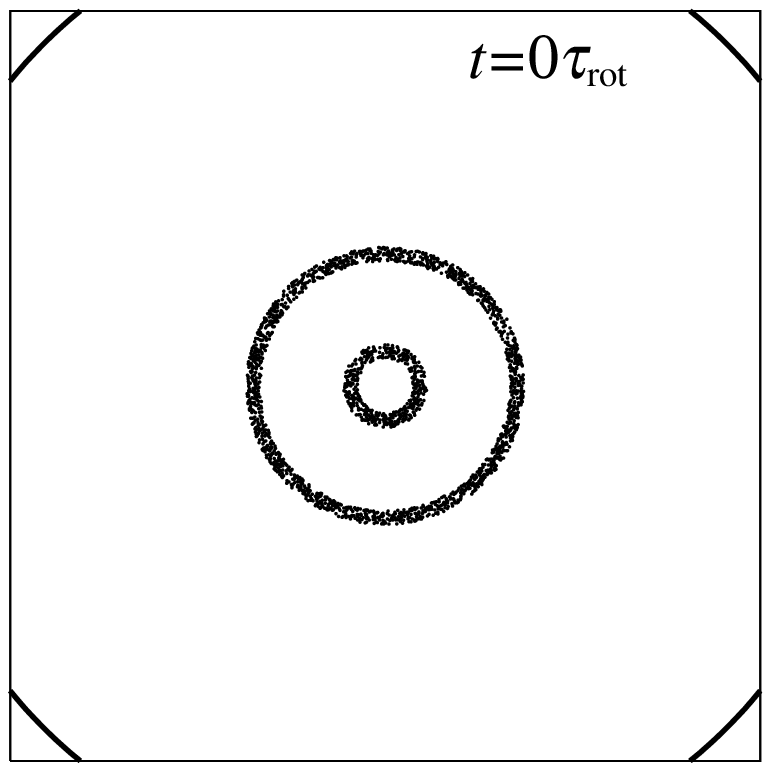}
   \includegraphics[trim=40 0 100 0,width=0.22\columnwidth]
                   {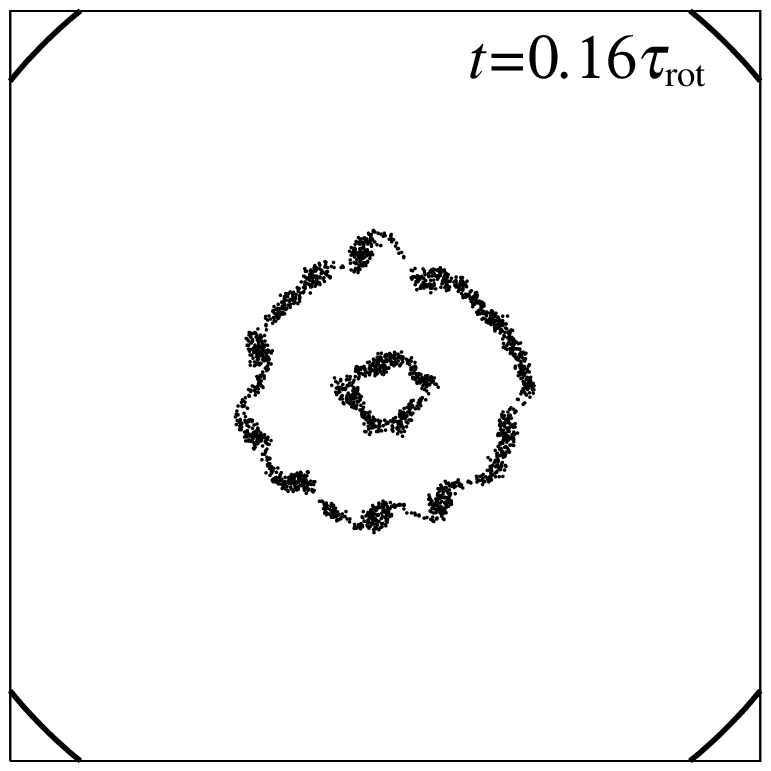}
   \includegraphics[trim=40 0 100 0,width=0.22\columnwidth]
                   {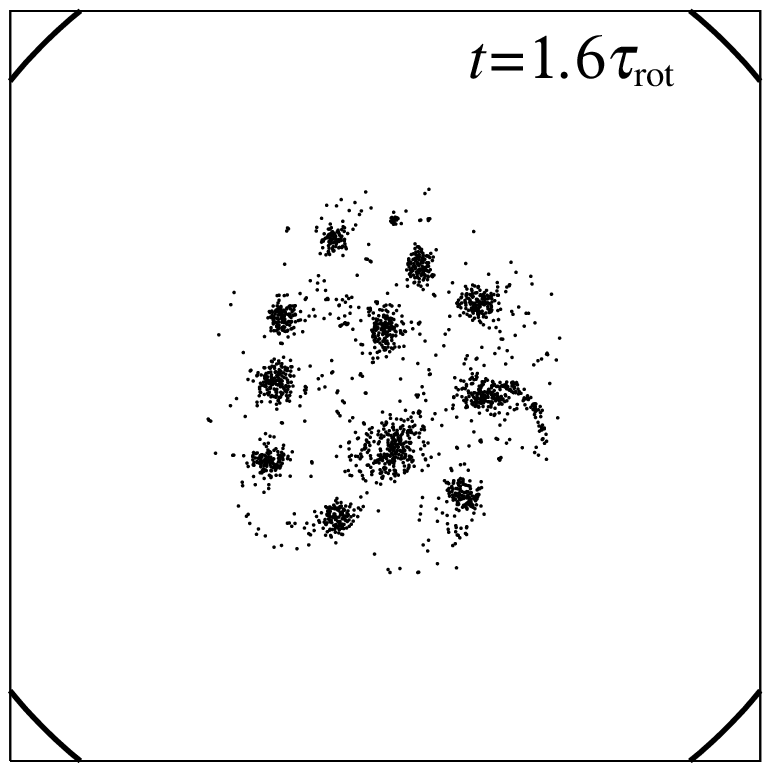}
   \includegraphics[trim=40 0 100 0,width=0.22\columnwidth]
                   {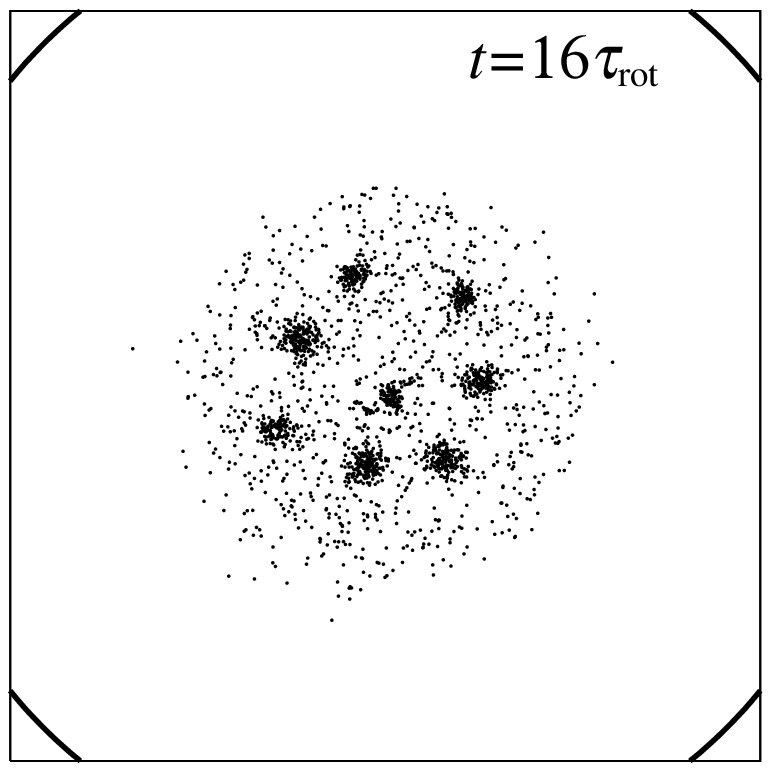}
  \caption{
   Time evolution of electron density distributions
   from
   (a) a single ring initial state, and
   (b) a double ring initial state.
   Initial distributions are shown in
   the left-most figures and quasi stationary states are
   shown in the right-most figures.
   The conductor wall boundary is 
   expressed by the solid arcs at the corner of the plotted area.
   }
  \label{f_typicaltest_n10srsp}
 \end{center} 
\end{figure}



In this subsection, we focus on the macroscopic behavior of the
particle density $n(\mib{r})$ obtained from coarse-grained observation
(From now on, we drop the subscript 'm' from the coarse-grained
field, $n_{\mathrm{m}}(\bvec{r})$ etc.).

It can be shown that the rotationally symmetric state with
the 
density $n(r)$ being a decreasing function of $r$ is not linearly unstable
and is a numerically stable
solution of eqs.  (\ref{m_euler}) $-$ (\ref{m_poisson}).
By contrast, the ring state,
where the electrons are distributed in a ring-shaped region,
is linearly unstable, and its instability is
called diocotron instability.
\cite{btheory_nonneutral_plasma}   

After the diocotron instability, nonlinear effects including
the complex stretching and folding of the density
structure take places.
These deformation causes a fine filament-like structure,
which will soon become too small to be observed in a finite resolution, 
and the system quickly approaches a quasi-stationary state.
\cite{
 bquasi_stationary_states
}

Figure \ref{f_typicaltest_n10srsp} shows two examples of the time
sequences,
which start from the two unstable initial states and lead to two types
of quasi-stationary states.
Figure \ref{f_typicaltest_n10srsp}(a) shows the sequence
that 
 starts from a single-ring
configuration,
 destabilizes into the mode three, and eventually falls
into a singly peaked stable density distribution.  In the case of
Fig. \ref{f_typicaltest_n10srsp}(b),
the instability of a double-ring initial configuration
leads to 
a higher mode to break into many vortices, which merge occasionally
while they undergo a collective chaotic motion, and eventually several
surviving vortices form a regular structure, which does not undergo
further change during our simulation time; such a state is
called a vortex crystal state.
\cite{btwo_dim_vortex_crystal}

These quasi-stationary states are in general not the equilibrium states
\cite{bquasi_stationary_states};
the quasi-stationary states strongly depend on the initial states,
which indicates that the system lacks the ergodicity
 in this time scale.
For longer time scale, these quasi-stationary states relax
into the equilibrium states, which we study
in detail in the followings.

\section{Statistical theories for diffusion and relaxation}
\label{s_theories}
In this section,
we briefly 
review some of the elements of statistical theory for
the slow relaxation.

%
\subsection{Maximum one-body entropy distribution for equilibrium state}
\label{s_theories_ss_equilibrium}
First, we describe the equilibrium state,
which should be given by the state that maximizes the 
entropy under some constraints
if the system is ergodic.

Within the one-body approximation,
\cite{
 bstatistics_line_vortices,
 bnegative_temperature_states,
 bnonaxisymmetric_thermal_equilibria
}
the 
entropy and the energy are approximated by the 
one-body entropy $S_{1}$ and the
mean field energy $H_{\mathrm{MF}}$
using  the coarse-grained density $n(\bvec{r})$ as,
\begin{equation}
 \label{m_point_vortex_entropy}
  S_{1} = -\int d^{2}\mib{r} \, n(\mib{r}) \log n(\mib{r}).
\end{equation}
\begin{equation}
 \label{m_meanfieldenergy}
  H_{\mathrm{MF}} 
    = -\frac{1}{2}\int d^{2}\mib{r}\int d^{2}\mib{r}' \,
        n(\mib{r})G(\mib{r}, \mib{r}')n(\mib{r}'),
\end{equation}
where $n(\bvec{r})$ is normalized to $N$ as
 in eq. (\ref{m_numberofparticles}).
The angular momentum $I$ is evaluated by 
substituting the coarse-grained density $n(\bvec{r})$
into eq. (\ref{m_angular_momentum}).
Then 
the theory becomes simple and the maximum one-body entropy state
is obtained
\cite{btwo_dim_turbulence_exp}
by maximizing eq. (\ref{m_point_vortex_entropy})
with respect to $n(\bvec{r})$ under the constraints of
$N$, $H_{\mathrm{MF}}$ and $I$ as
\begin{equation}
 n(\mib{r})
  = \exp(-1 - a + b \phi(\mib{r}) - c r^{2}).
\end{equation}
with the Lagrange multipliers $a, b,$ and $c$.
From  eq. (\ref{m_poisson}), $\phi$ satisfies
\begin{equation}
 \nabla^{2}\phi(\mib{r})
  = \exp(-1 - a + b \phi(\mib{r}) - c r^{2}),
\end{equation}
which can be solved numerically .

Note that $S_{1}$ would be a Casimir constant if
the density field strictly followed the Euler equation (\ref{m_euler}).

\subsection{Velocity distribution of randomly placed point vortex system}
\label{s_theories_ss_velocitydistribution}
The velocity distribution of particles is an
important element in the kinetic theory.
In an ordinary system, the velocities and the positions of
particles are two sets of dynamical variables that define
a state. In the present system of the point vortices, however,
the velocities are determined by the positions, and
its distributions has been analyzed in terms of
Levy's stable distribution.
\cite{
 bstable_distribution_levi,
 blevy_stable_distributions,
 bholtsmark_distributions,
 bstochastic_problems_physics_astronomy}



Let us consider the pancake density distribution,
where the particles are located randomly with
a constant average density  within a radius $R$.
The system undergoes a rigid bulk rotation with the period
$\tau_{\mathrm{rot}}$ on average.
At the center of the system, the particles have
no average velocity and only the velocity fluctuation exists.
For large $N$,
the velocity distribution at the center of the system
has been found to be the Gaussian
with the variance proportional to $n\log N$
 for the small velocity region, 
and the power-law $V^{-4}$  in the large velocity region,
\cite{bstatmech_twodim_vorices_stellar}
\begin{equation}
\label{mveldist}
W(\bvec{V}) = 
\left\{
\begin{array}{ll}
 \frac{4}{n\gamma^{2}\log N}
 \exp(-\frac{4\pi}{n\gamma^{2}\log N}V^{2}),
 &
 (V \ll V_{\mathrm{crit}}(N)) \\
 \frac{n\gamma^{2}}{4\pi^{2}V^{4}},
 &
  (V \gg V_{\mathrm{crit}}(N)),
\end{array}
\right.
\end{equation}
with the crossover velocity
\begin{equation}
 V_{\mathrm{crit}}(N)
  \sim
 \left(\frac{n\gamma^{2}}{4\pi}\log N\right)^{1/2}
 [\log(\log N)]^{1/2} ,
\end{equation}
where $\gamma$ is the circulation (or electric charge) of a particle
and normalized to unity in this paper.
Note that the power law
behavior of the distribution in
large $V$ comes from the distance distribution of the 
closest particles among the randomly located particles.


\subsection{Kinetic theory of point vortex system}
\label{s_theories_ss_kinetic}
Exploiting the similarity to the relaxation process
of the self-gravitating system,
\cite{bstatistical_mechanics_gravitating}
Chavanis has developed the kinetic theory for
the point vortex system.
\cite{
 bkinetic_theory_of_point_vortices,
 bstatmech_twodim_vorices_stellar}
As we have described, there are two relaxations,
i.e. the fast relaxation and the slow relaxation;
we will briefly review some of his results
for the slow relaxation.

After the system establishes a quasi-stationary state
following the Euler equation, small effects of the
collision term 
 in eq. (\ref{m_euler_collision}) gradually stir the system
to cause the slow relaxation, or the collisional relaxation.
Chavanis
has estimated
the time scale of the collisional
 relaxation  in terms of
the diffusion of point vortices caused by the collision term
as follows.

Assume that the velocity distribution is the Gaussian
given in eq. (\ref{mveldist})
for the whole $V$ region,
then we can estimate the typical velocity $V_{\mathrm{typ}}$ from
 the mean square
velocity as,
\begin{equation}
 V_{\mathrm{typ}}^{2}
 \equiv
 \langle V^{2} \rangle = \frac{n\gamma^{2}}{4\pi}\log N .
\end{equation}

For the pancake state with the flat distribution,
there is no shear flow, thus the diffusion
constant $D$ may be estimated as
\begin{equation}
 D \sim l_{0}V_{\mathrm{typ}} \sim \gamma \sqrt{\log N}
\end{equation}
where $l_{0} \sim 1 / \sqrt{n_{0}}$ is the average inter-particle
distance.


For the state with singly peaked distribution,
there exists the shear flow around the
peak, therefore, the diffusion is not isotropic.
The diffusion coefficient for the radial direction may be estimated as
\begin{equation}
 D \sim \tau V_{\mathrm{typ}}^{2}
 \sim
 \frac{\gamma n(\bvec{r})}{|\Sigma(\bvec{r})|} \log N
\end{equation}
where the correlation time $\tau$ has been estimated as 
$\tau \sim 1 / |\Sigma|$ with the local shear 
\begin{equation}
 |\Sigma(\bvec{r})| = r\frac{d}{dr}\left(\frac{V(r)}{r} \right) .
\end{equation}

If the slow relaxation is due to this diffusion process, the relaxation
time $\tau_{\mathrm{relax}}$ would be the time for a particle
to diffuse over the system size:
\begin{equation}
 \tau_{\mathrm{relax}} \sim \frac{L^{2}}{D}
 \sim \frac{N}{\sqrt{\log N}}\tau_{\mathrm{rot}}
\end{equation}
for the pancake state without shear flow, and
\begin{equation}
 \tau_{\mathrm{relax}} \sim \frac{L^{2}}{D}
 \sim \frac{N}{\log N}\tau_{\mathrm{rot}}
\end{equation}
for the singly peaked  state with shear flow.
In the both cases, $\tau_{\mathrm{relax}}$
increases almost linearly with $N$ in units of the bulk rotation
time.



\section{Simulation method}
\label{ss_method}
We simulate the point vortex system by integrating
eqs. (\ref{m_hamiltonian_eq}) $-$ (\ref{mcylinder_potential})
\cite{bvortex_method_flow_simulations}.
Number of particles $N$ ranges over $N = 128 \sim 2048$.
Force acting on each particle is calculated by
the simple sum of Coulomb force from
all the other particles, therefore
computational complexity
 of calculating the interactions is $O(N^{2})$ for
each time step.
Numerical integration in time is performed
 using the second-order Runge-Kutta method.
The time step of the integration is
 $\Delta t = 0.005\tau_{\mathrm{rot}}/4\pi$.
The  error $\Delta H$ in the total energy $H$
tends to increase almost linearly
in time, and the average increase of the error is
$\Delta H / H \sim 10^{-5}$ for $N = 128$ and
$\Delta H / H \sim 10^{-7}$ for $N = 2048$
per bulk rotation period $\tau_{\mathrm{rot}}$.

Initial configurations of particles are generated using 
random number sequences to be consistent with a given macroscopic
density distribution.
The problem is that,
if two particles are too close to each other,
they rotate around each other
with large velocity; This could cause large integration
error. To avoid this problem, we
restrict the distance between any pair of particles to be larger
than a certain limit length proportional to $\sqrt{\Delta t/ N}$
in an initial configuration.
This restriction is implemented by
re-generating a particle position if 
the newly placed particle is
too close to the existing particles.
It should be noted, however, that
 such initial condition does not ensure
that the particles do not come close to each other
in future.

The unit length is defined so that the
average square distance $\lambda^{2}$ from the center
to be
\begin{equation}
 \label{m_unitlength}
 \lambda^{2} \equiv \frac{\sum_{i}^{N}r_{i}^{2}}{N} = \frac{1}{2},
\end{equation}
which is the angular momentum
(\ref{m_angular_momentum}) per particle,
hence, is a constant of dynamics.
The radius of the container  $R_{\mathrm{w}}$ is chosen
as
\begin{equation}
  R_{\mathrm{w}} = 2.9,
\end{equation}
which is comparable to that in the experiments.
\cite{btwo_dim_turbulence_exp, btwo_dim_vortex_crystal}
With this value, the container
 does not impose any conceivable restrictions on
the relaxation process.



\section{Simulation results}
\label{s_simulation}

In this section we present our simulation results,
mainly on the slow relaxation process
in comparison with the kinetic theory by Chavanis.

After describing a couple of system setups used in
the numerical
simulation (\S\ref{s_simulations_ss_initial}),
we present general behavior
of the system and show that the time scale of the
slow relaxation grows almost linearly with
$N$ (\S\ref{s_simulations_ss_results_sss_slowrelax}).
The velocity distributions are shown in 
\S\ref{s_simulations_ss_results_sss_velocitydistribution}.
Particle trajectories are examined in
\S\ref{s_simulations_ss_results_sss_anomalousdiffusion}
to find anomalous diffusion, which is
analyzed in terms of the Levy flight by decomposing
the trajectories into sequences of steps,
whose length and duration distributions are found to
be of the power laws (\S\ref{s_simulations_ss_results_sss_randomwalk}).

%
%

\subsection{Initial configurations}
\label{s_simulations_ss_initial}

We will present simulation data mainly for 
the system setups that starts from
two different initial states, which we call
Setup A and Setup B in the following.

Setup A is the simulation that starts from
 the pancake state
with a constant macroscopic density distribution in the
region $r < R_{0}$,
where the radius of the outer edge of the pancake $R_{0}$ 
 is determined by the condition
(\ref{m_unitlength}).
In the continuum limit, this pancake state
is the minimum energy state for a given angular momentum,
 therefore, the system cannot relax macroscopically.

Setup B is the simulation that starts from a 
double ring distribution.
The particles are distributed in the regions
$0.4R_{0} < r < 0.6R_{0}$ and $0.8R_{0} < r < R_{0}$.
In this case, the system relaxes into a singly peaked state
with shear flow,
as we will see below.

The numbers of particles $N$ are 
same for both setups A and B.
The relaxation properties of the macroscopic density
 are mainly obtained from 
Setup B of the double ring initial state.
Setup A of the pancake initial state
 is used to investigate the diffusion process
of individual point vortices.

Another
variation of initial configuration, which we call
Setup B', is used
in Fig. \ref{f_fn_comp_rdavg} in order to see
the initial state dependence of the slow relaxation.
Setup B' is the simulation that starts from a 
single ring distribution.
The particles are distributed in the regions
$0.4R_{0} < r < R_{0}$.
The energy happens to be almost the same with that of Setup B.
The system relaxes into a singly peaked state.

\subsection{Fast and slow relaxation}
\label{s_simulations_ss_results_sss_slowrelax}
\begin{figure}[b]
 \begin{center}
   \includegraphics[trim=20 0 0 0,width=0.40\columnwidth]
                   {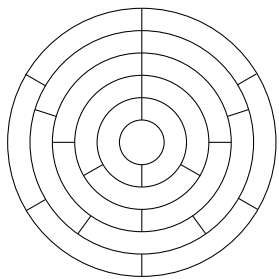}
  \caption{
    An example of mesh used to observe the macroscopic density field
    for one-body entropy calculation.
    Number of partition in azimuthal direction increases by one 
    as $r$ increases
    (in the figure, $1, 2$ and $3$ from the center to the periphery).
    In the actual  observations, the resolution in
    radial direction $m_{r}$ is $m_{r} = 20$.
  }
  \label{f_entropy_mesh}
 \end{center}
\end{figure}

\begin{figure}[b]
 \begin{center}
   \includegraphics[trim=20 0 80 0,width=0.42\columnwidth]
                   {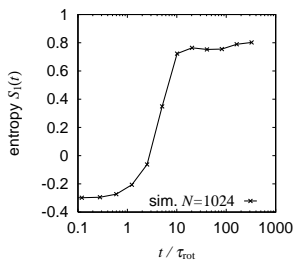}
  \caption{
    Time development of the one-body entropy $S_{1}(t)$ in the
    semilogarithmic
    scale for Setup B with $N = 1024$.
  }
  \label{f_fastslowrelax}
 \end{center}
\end{figure}

Figure \ref{f_fastslowrelax} shows the time development of the 
one-body entropy $S_{1}(t)$ as a function of time $t$ in 
the semilogarithmic scale.
The one-body entropy is calculated by coarse-graining the 
distribution of point vortices using a mesh shown in
Fig. \ref{f_entropy_mesh}.
The size of the cell at distance $r$ from the center
is $\delta r = R_{\mathrm{w}}/m_{r}$ in 
the radial direction
 and is $r\delta \theta = 2\pi R_{\mathrm{w}} /  m_{r}$
in the azimuthal direction, where $m_{r}$ is the number of 
division in the radial direction, thus the shape of a cell is
elongated in the azimuthal direction.
This gives the area of a cell
 $\delta A \approx (\delta r)(r\delta \theta) = 2A/m_{r}^{2}$
where $A = \pi  R_{\mathrm{w}}^{2}$ is the total area of the system.
Here, $m_{r} = 20$ is used for all the simulations.

We observe three stages in the development of $S_{1}(t)$:
 (i) diocotron instability ($t \lesssim 1\tau_{\mathrm{rot}}$),
     where $S_{1}(t)$ is almost constant,
(ii) fast relaxation
     ($1\tau_{\mathrm{rot}} \lesssim t \lesssim 10\tau_{\mathrm{rot}}$),
     where $S_{1}(t)$ shows a rapid increase,
 and
(iii) slow relaxation ($10\tau_{\mathrm{rot}} \lesssim t $),
     where $S_{1}(t)$ increases slowly with time.

\begin{figure}[b]
 \begin{center}
   \includegraphics[trim=20 0 80 0,width=0.42\columnwidth]
                   {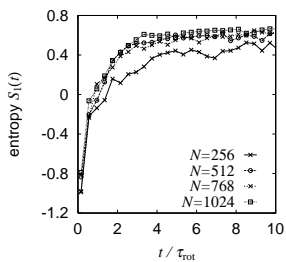}
  \caption{
    Time evolution of the one-body entropy $S_{1}$ during
    the fast relaxation
    for various $N$ for Setup B.
    The time $t$ is scaled by the bulk rotation time 
    $\tau_{\mathrm{rot}} \propto 1/N$.
  }
  \label{f_fastrelax}
 \end{center}
\end{figure}

The one-body entropy $S_{1}(t)$ as a function of time 
$t$ in the fast relaxation is shown in Fig. \ref{f_fastrelax}
for various $N$.
We do not observe the systematic $N$-dependence of 
the relaxation time scale
if we scale the time $t$ by the bulk rotation period
$\tau_{\mathrm{rot}} \propto 1/N$.

\begin{figure}[b]
 \begin{center}
  \raisebox{0.36\columnwidth}{(a)}
  \hspace{-1em}
  \includegraphics[trim=20 0 80 0,width=0.42\columnwidth]
                  {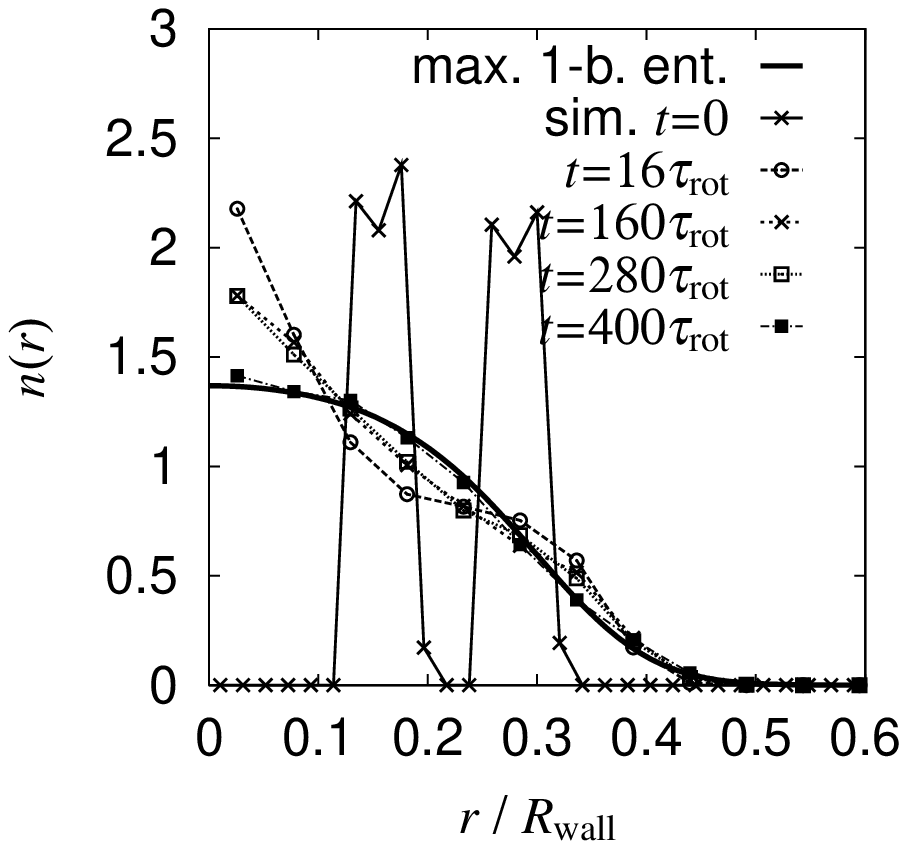}
  \raisebox{0.36\columnwidth}{(b)}
  \hspace{-1em}
  \includegraphics[trim=20 0 80 0,width=0.42\columnwidth]
                  {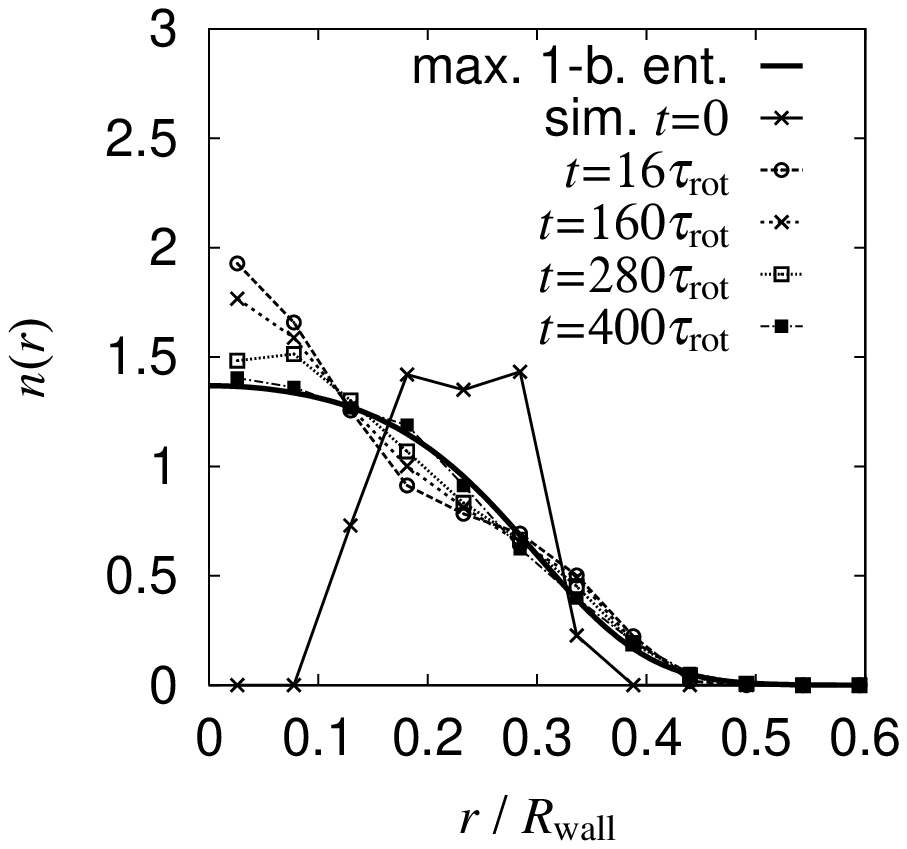}
  \caption{
    Time development of the density distribution 
    vs. $r$ for 
    the double ring initial state (Setup B) (a), and
    the single ring initial state (Setup B') (b).
    In the both cases, the number of particles is  $N=1024$ and
    the energy are almost the same for the both cases.
    The maximum one-body entropy state is also
    shown by the thick solid curves.
    For each curves of simulation results,
    the density distribution is averaged over 100 samplings
    at different times within the duration of about 10 rotations.
  }
  \label{f_fn_comp_rdavg}
 \end{center}
\end{figure}

\begin{figure}[b]
 \begin{center}
   \includegraphics[trim=40 0 100 0,width=0.22\columnwidth]
                   {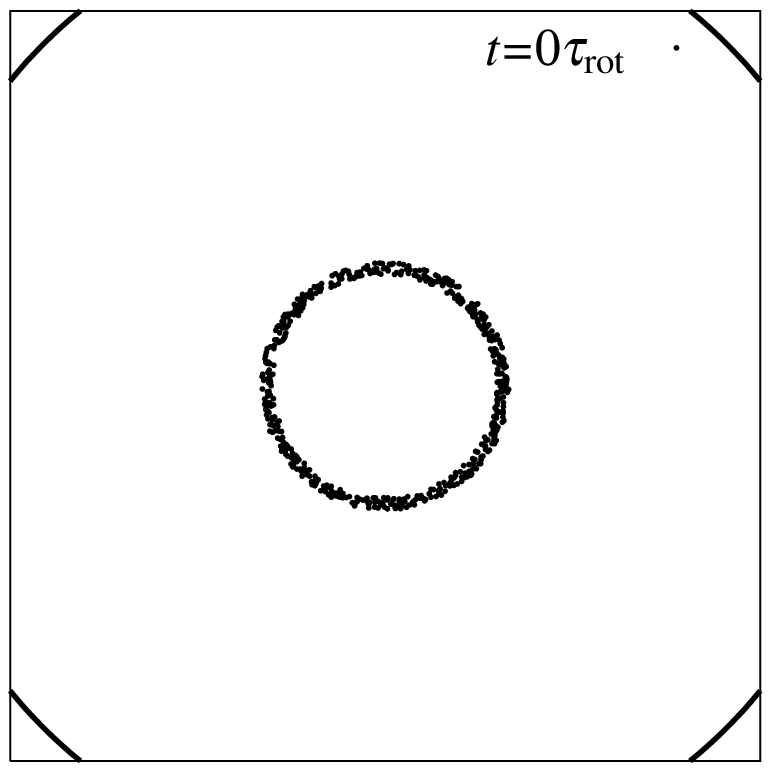}
   \includegraphics[trim=40 0 100 0,width=0.22\columnwidth]
                   {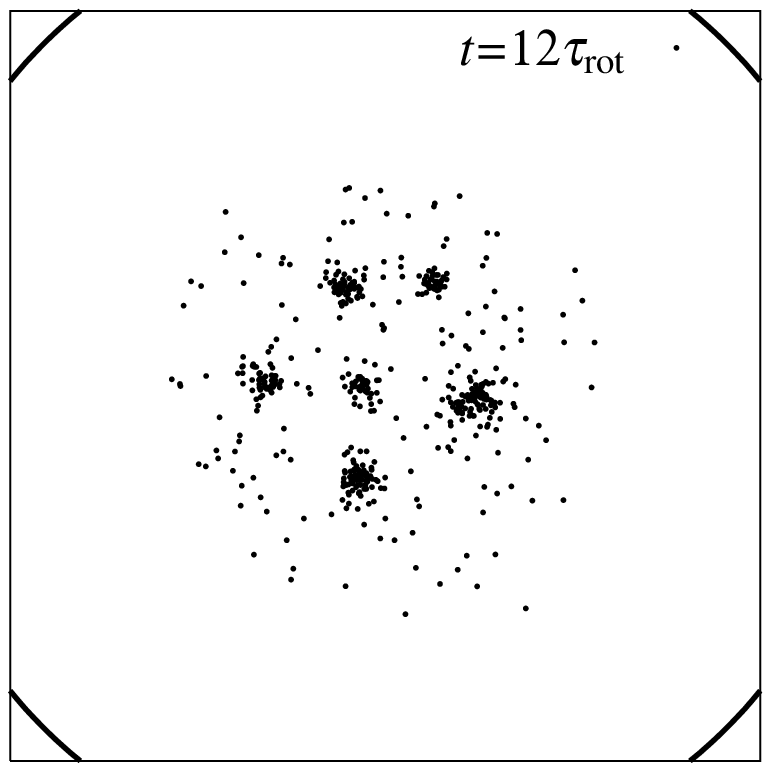}
   \includegraphics[trim=40 0 100 0,width=0.22\columnwidth]
                   {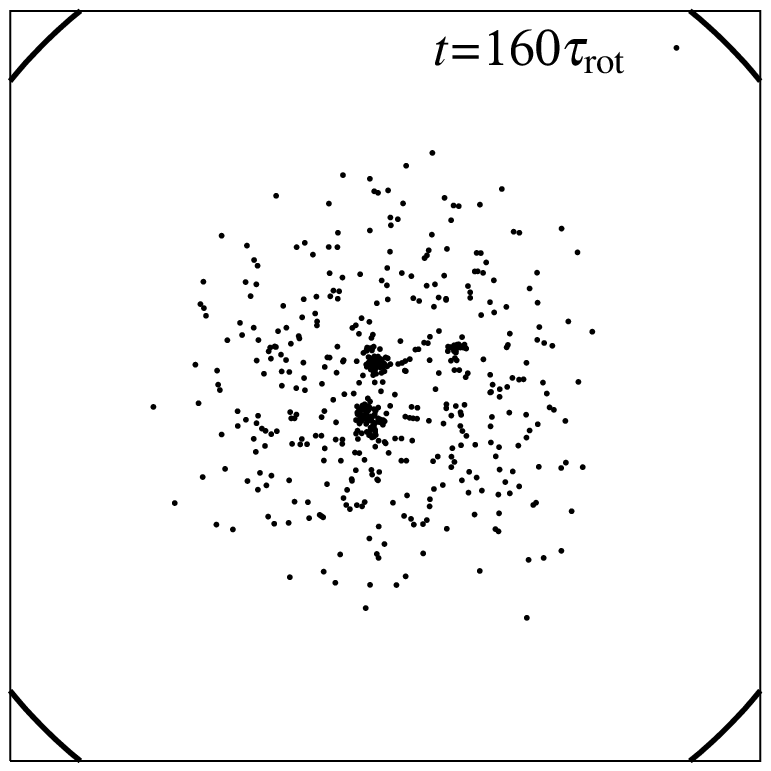}
   \includegraphics[trim=40 0 100 0,width=0.22\columnwidth]
                   {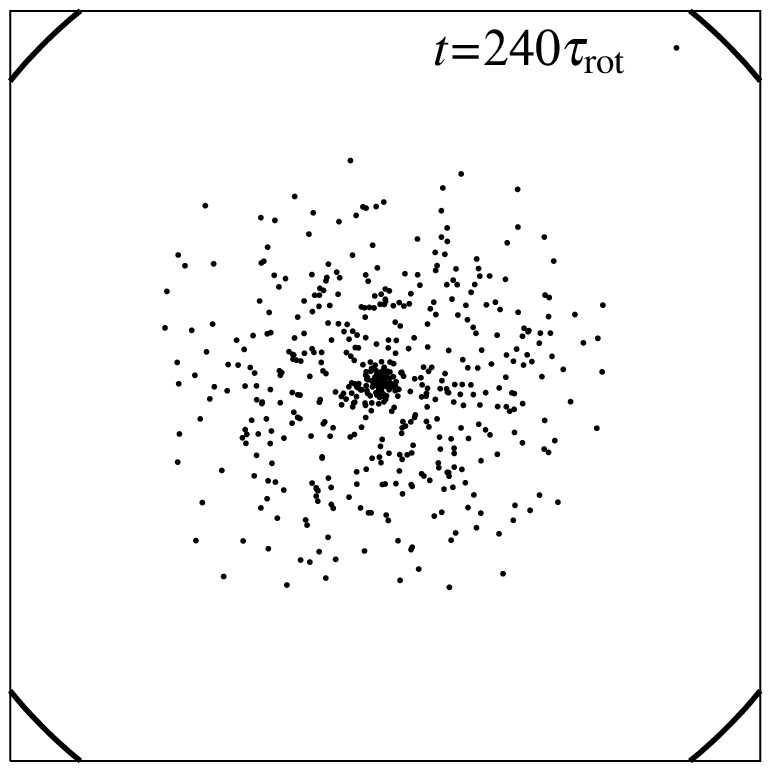}
  \caption{
    Time evolution of the point vortex distribution with
    $N = 512$.
    The initial state is the single ring state with a
    constant density at 
    $0.9R_{0} < r < R_{0}$ of randomly placed particles.
   }
  \label{f_vccollapse}
 \end{center} 
\end{figure}

Figure \ref{f_fn_comp_rdavg}
shows the time evolutions of 
the density distributions as a function of $r$.
The initial states are
the double ring initial state (Setup B) (a),
and 
the single ring initial state (Setup B') (b).
The maximum one-body entropy state is also shown
by the thick solid lines for comparison.
The energy and the angular momentum are same for these two cases,
so that the maximum one-body entropy state is same
for the both cases.
Although the quasi-stationary state achieved after the fast relaxation
depends on the initial state,
after the slow relaxation, the density distribution
approaches the maximum one-body entropy state in both cases.
Another example is shown in Fig. \ref{f_vccollapse},
where the initial single ring state relaxes into
 the vortex crystal state after
the fast relaxation,
but the vortices are smeared to merge into a single broad peak
and the state eventually approaches  the maximum one-body entropy
state through the slow relaxation.

\begin{figure}[b]
 \begin{center}
   \includegraphics[trim=20 0 80 0,width=0.42\columnwidth]
                   {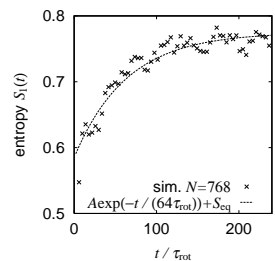}
  \caption{
    Time evolution of the one-body entropy $S_{1}$
    during the  slow relaxation for
    Setup B
    with $N = 768$.
    The simulation data are averaged over 100 successive points.
    The dashed curve shows an exponential decay of
    eq. (\ref{m_exp_entropy}),
    with the fitting parameters
    $S_{\mathrm{eq}} = 0.775$,
    $A = 0.19$,
    and $\tau_{\mathrm{relax}} = 64\tau_{\mathrm{rot}}$.
  }
  \label{f_exp_relax}
 \end{center}
\end{figure}
\begin{figure}[b]
 \begin{center}
   \includegraphics[trim=20 0 80 0,width=0.42\columnwidth]
                   {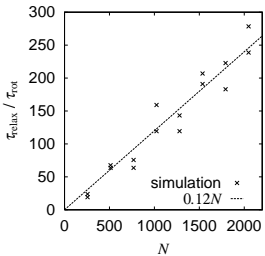}
  \caption{
   Relaxation time $\tau_{\mathrm{relax}}$ of the one
   body entropy in the slow relaxation as a function of
   the number of particles $N$.
   The $y$-axis is normalized by the bulk rotation time
   $\tau_{\mathrm{rot}}$.
   The simulation is by Setup B.
  }
  \label{f_comp_relaxtime}
 \end{center}
\end{figure}
Figure \ref{f_exp_relax} shows that
the time development of the one-body entropy $S_{1}(t)$ during the slow
relaxation can be approximated by the exponential form:
\begin{equation}
\label{m_exp_entropy}
 S(t) \approx S_{\mathrm{eq}} - A\exp(-t/\tau_{\mathrm{relax}}).
\end{equation}
We determine the parameters $S_{\mathrm{eq}}, A$ and
$\tau_{\mathrm{relax}}$ by fitting with the simulation data.
The  relaxation time scale
$\tau_{\mathrm{relax}}$ 
is plotted against the number of particles $N$
 in Fig. \ref{f_comp_relaxtime}.
We find that $\tau_{\mathrm{relax}}$ increases almost linearly with
$N$ if $\tau_{\mathrm{relax}}$ is measured in the unit
 of the bulk rotation time
$\tau_{\mathrm{rot}} \propto 1/N$. This is consistent with the 
theory of Chavanis  in \S\ref{s_theories_ss_kinetic}.

\subsection{Velocity distribution}
\label{s_simulations_ss_results_sss_velocitydistribution}

\begin{figure}[b]
 \begin{center}
   \raisebox{0.36\columnwidth}{(a)}
   \hspace{-1em}
   \includegraphics[trim=50 0 50 0,width=0.42\columnwidth]
                   {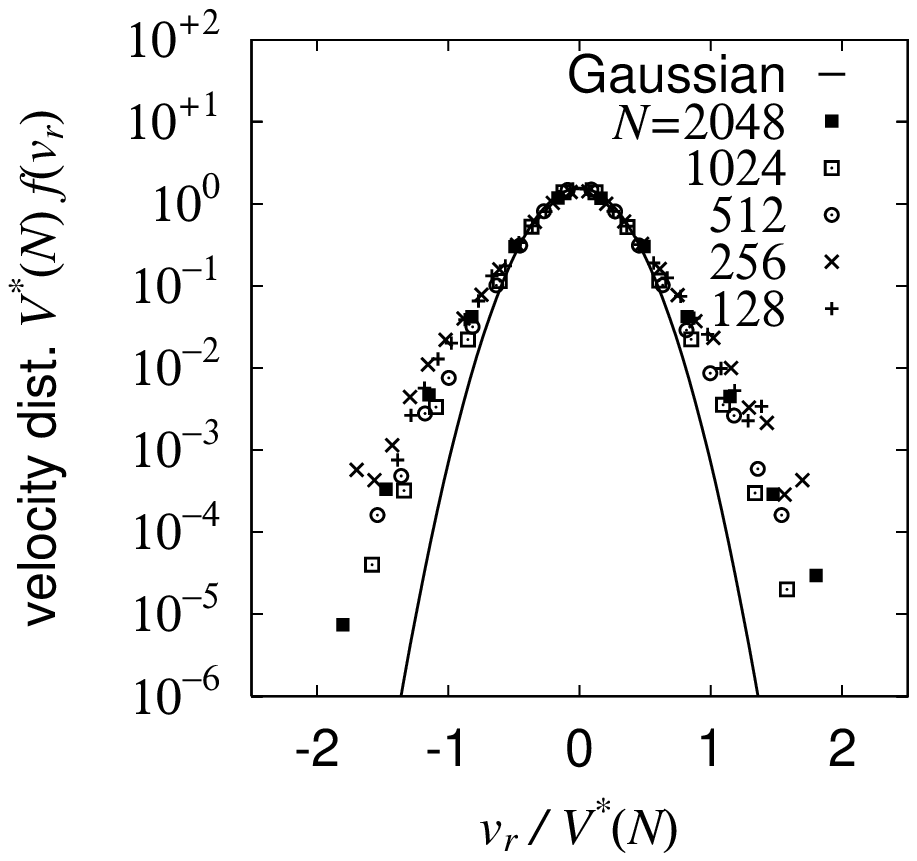}
   \raisebox{0.36\columnwidth}{(b)}
   \hspace{-1em}
   \includegraphics[trim=50 0 50 0,width=0.42\columnwidth]
                   {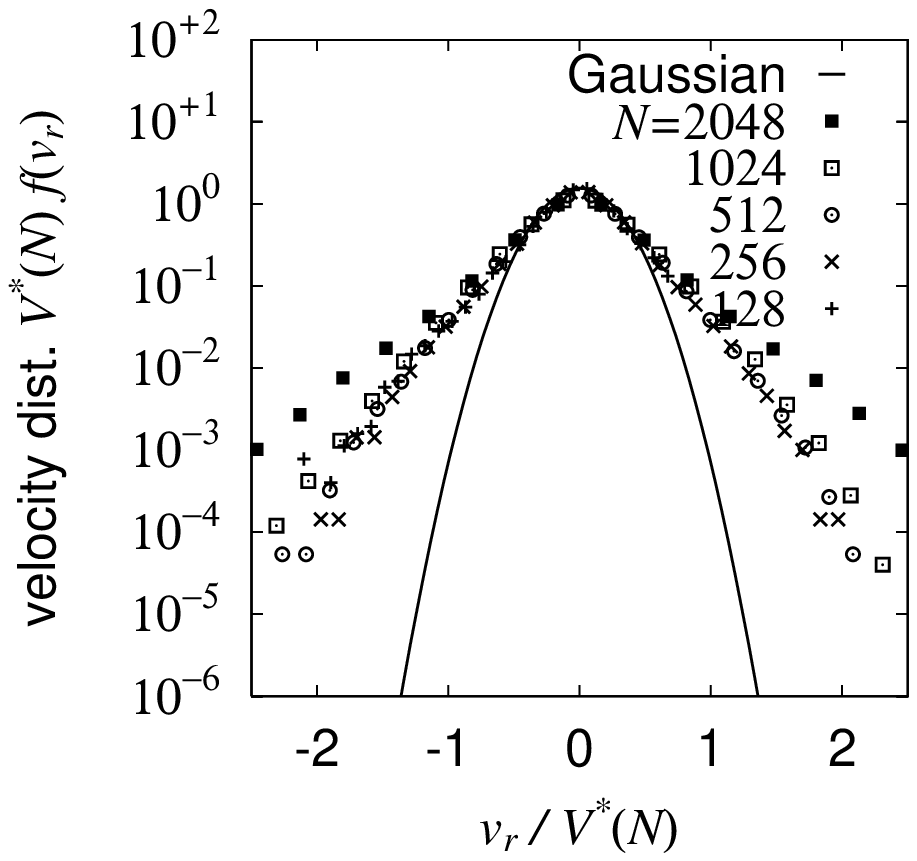}
  \caption{
    Distributions of the radial component of velocity
    of two different simulation setups:
    Setup A (the pancake state) (a), and
    Setup B (singly peaked state) (b).
    The velocity is scaled by $V^{*}(N) \equiv N\log N$.
    The distribution functions are averaged over 200 samplings
    at different times during the slow relaxation,
    $80\tau_{\mathrm{rot}} \leq t \leq 240\tau_{\mathrm{rot}}$.
  }
  \label{f_ps_avgvsar-scaled}
 \end{center}
\end{figure}

\begin{figure}[b]
 \begin{center}
   \includegraphics[trim=20 0 80 0,width=0.42\columnwidth]
                   {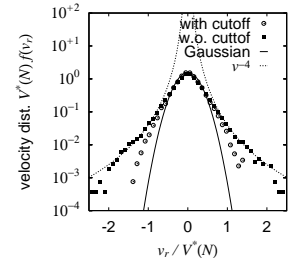}
  \caption{
    Distributions of the radial component of velocity
    for
    Setup A with ($\bigcirc$) and without ($\blacksquare$)
    the restriction on the distance between the particles
    (see \S\ref{ss_method}).
    The number of particles is  $N = 128$.
    The velocity is scaled by $V^{*}(N) \equiv N\log N$.
    The distribution functions are averaged over 200 samplings at
    different times during the slow relaxation,
    $80\tau_{\mathrm{rot}} \leq t \leq 240\tau_{\mathrm{rot}}$.
  }
  \label{f_ps_factor0-avgvsar-scaled}
 \end{center}
\end{figure}

Figure  \ref{f_ps_avgvsar-scaled} shows
the radial velocity distributions
of 
the simulations of
Setup A (the pancake state) (a), and
Setup B (the singly peaked state)(b).
In both cases, we measure
the velocity distribution in the same time region
$80\tau_{\mathrm{rot}} \leq t \leq 240\tau_{\mathrm{rot}}$,
which 
is within the slow relaxation for the case of
the singly peaked state (b),
while the pancake state (a) does not show
any macroscopic relaxation.
Only the distributions of the radial component
of velocity are plotted
because the azimuthal velocity contains the average bulk rotation.


In Fig. \ref{f_ps_avgvsar-scaled}(a), the Gaussian distribution
that scales as $v / (N\log N)$ is observed for small $v$,
as is expected by the theory in \S\ref{s_theories_ss_kinetic}.
\cite{
 bstable_distribution_levi,
 blevy_stable_distributions,
 bholtsmark_distributions,
 bstatmech_twodim_vorices_stellar}
For large $v$, however, we find the exponential tail
 instead of the power law tail.

The reason of the lack of the power law tail seems to be
the restriction of the particle distance
imposed on the initial state,
as is explained in \S\ref{ss_method}.
If we remove this restriction and locate the particles
randomly, then we obtain the tail closer to the power law
$v^{-4}$, as shown in
Fig. \ref{f_ps_factor0-avgvsar-scaled}.
Note that the precision of the energy conservation is poor
for the simulation without the  restriction.

For the singly peaked state (Fig. \ref{f_ps_avgvsar-scaled}(b)),
we find that the distribution is broader than that for
the pancake case (a) but the tail is still exponential.
Although the distribution does not fit
to the Gaussian very well, it
is still scaled by $v / (N\log N)$ in small $v$ region.

\subsection{Anomalous diffusion}
\label{s_simulations_ss_results_sss_anomalousdiffusion}

\begin{figure}[b]
 \begin{center}
   \includegraphics[trim=20 0 80 0,width=0.42\columnwidth]
                   {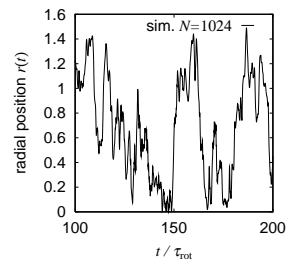}
  \caption{
    Time development of the radial position of a particle
    for  Setup B
    (the singly peaked state) with $N = 1024$.
  }
  \label{f_randomwalk}
 \end{center}
\end{figure}

Figure \ref{f_randomwalk} shows a trajectory of the
 radial position of a particle in the quasi-stationary state.
One can see that the motion of a particle looks like a random walk,
while it occasionally undergoes long jumps with a step
 almost of the system size
($\Delta r \sim 1$).

\begin{figure}[b]
 \begin{center}
   \raisebox{0.36\columnwidth}{(a)}
   \hspace{-1em}
   \includegraphics[trim=50 0 50 0,width=0.42\columnwidth]
                   {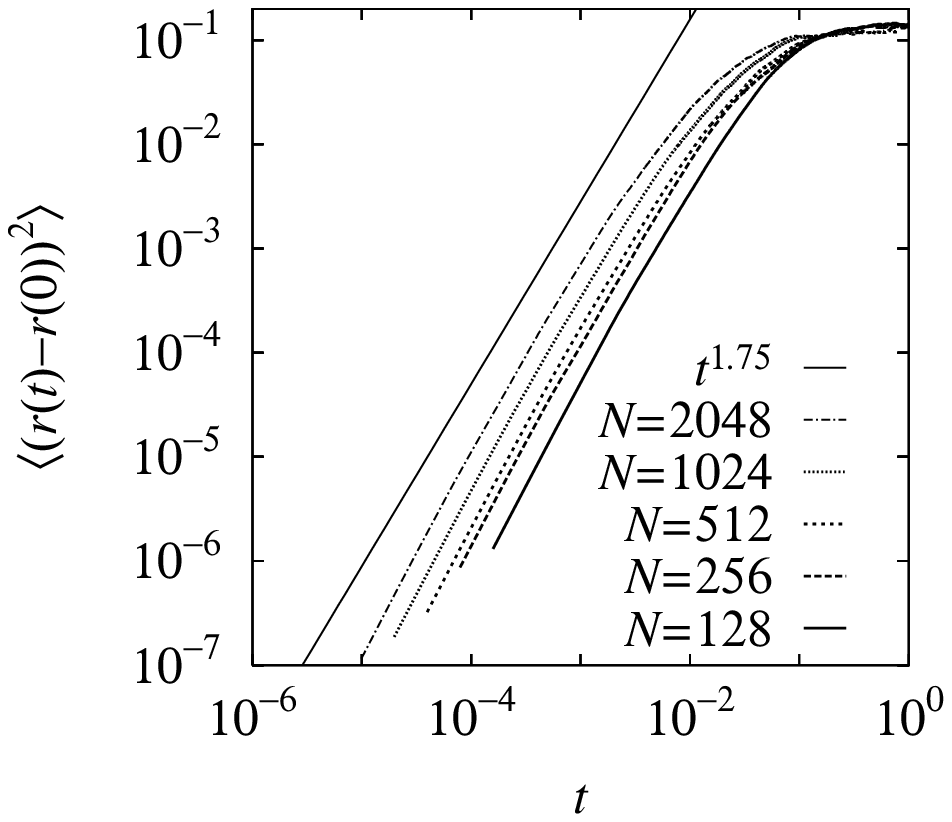}
   \raisebox{0.36\columnwidth}{(b)}
   \hspace{-1em}
   \includegraphics[trim=50 0 50 0,width=0.42\columnwidth]
                   {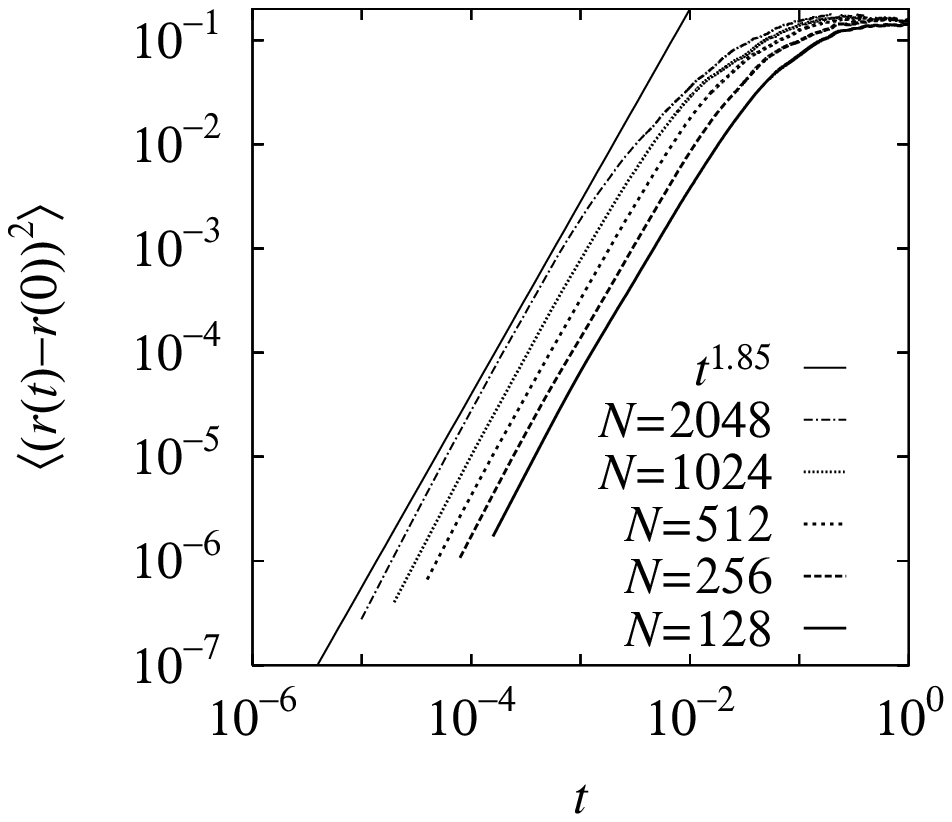}
  \caption{
    Time evolution of the mean square of the difference of radial
    position $\Delta r(t) = r(t) - r(0)$
    from its initial position in the log-log plot.
    The simulation setups are
    Setup A (the pancake state) (a), and
    Setup B (the singly peaked state) (b).
  }
  \label{f_ps_anomalous_avgdev}
 \end{center}
\end{figure}

To see the diffusion behavior, the time development
of the mean square radial deviation $\langle (\Delta r(t))^{2} \rangle$
where $\Delta r(t) = r(t) - r(0)$
is plotted in Fig. \ref{f_ps_anomalous_avgdev}(a) for the pancake case
with various $N$ (Setup A).
We observe a superdiffusion
\begin{equation}
\label{m_anomalous_diff}
 \langle (\Delta r(t))^{2} \rangle \sim t^{\gamma},
\end{equation}
with the diffusion exponent $\gamma = 1.75 \pm 0.1 > 1$
for all $N$.
The square radial deviation saturates in
 $\langle (\Delta r(t))^{2} \rangle \approx 0.1$
due to the finite system size $L \sim 1$.
From Fig. \ref{f_ps_anomalous_avgdev}(b), we obtain
$\gamma = 1.85 \pm 0.1$ for the case of Setup B (the singly peaked
state).

\begin{figure}[b]
 \begin{center}
   \includegraphics[trim=50 0 50 0,width=0.42\columnwidth]
                   {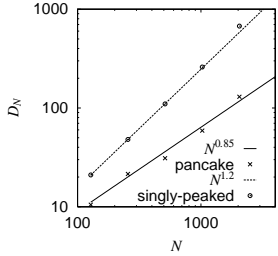}
  \caption{
    Coefficient of anomalous diffusion
    $D_{N} = \langle \Delta r^{2}\rangle / t^{\gamma}$
    as a function of the number of particles $N$.
    The coefficient $D_{N}$ are estimated from the plots in
    Fig. \ref{f_ps_anomalous_avgdev}.
    $\gamma = 1.75$ is used for 
    Setup A (the pancake state), and
    $\gamma = 1.85$ is used for 
    Setup B (the singly peaked state).
  }
  \label{f_ps_coeff_anomalous_diff}
 \end{center}
\end{figure}

Figure \ref{f_ps_coeff_anomalous_diff} shows
the ``coefficient of anomalous diffusion'' defined as
\begin{equation}
 D_{N} = \frac{\langle (\Delta r(t))^{2} \rangle}{t^{\gamma}},
\end{equation}
 as
a function of $N$.
We find a power law dependence on $N$ as
$D_{N} \sim N^{\eta}$ with $\eta = 0.85 \pm 0.1$.

\begin{figure}[b]
 \begin{center}
   \includegraphics[trim=50 0 50 0,width=0.42\columnwidth]
                   {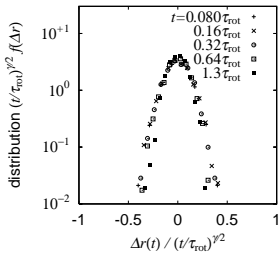}
  \caption{
   Distribution $P(\Delta r|t)$ of the deviation of the radial position
   $\Delta r(t) = r(t) - r(0)$ at various times $t$.
   The plotted data are same with those
   in Fig.  \ref{f_ps_anomalous_avgdev}(a) for $N = 2048$.
   The deviation $\Delta r$ is scaled by $t^{\gamma / 2}$
   with the same exponent $\gamma = 1.75$ with
   Fig. \ref{f_ps_anomalous_avgdev}(a).
  }
  \label{f_ps_xtscaletest}
 \end{center}
\end{figure}

Figure \ref{f_ps_xtscaletest} shows the distribution function
$P(\Delta r | t)$ of the deviation $\Delta r$ at various times $t$.
The data are taken from the same simulation with that
 in Fig. \ref{f_ps_anomalous_avgdev} (a) for $N = 2048$.
We see that the distribution curves overlap by the scaling,
$\Delta r / t^{\gamma / 2}$ with $\gamma = 1.75$.
which shows the width of the distribution grows as
$t^{\gamma / 2}$.
The tail of the distribution converges to zero faster than the
exponential.

\subsection{Random walk  and Levy flight}
\label{s_simulations_ss_results_sss_randomwalk}

\begin{figure}[b]
 \begin{center}
   \raisebox{0.36\columnwidth}{(a)}
   \hspace{-1em}
   \includegraphics[trim=50 0 50 0,width=0.42\columnwidth]
                   {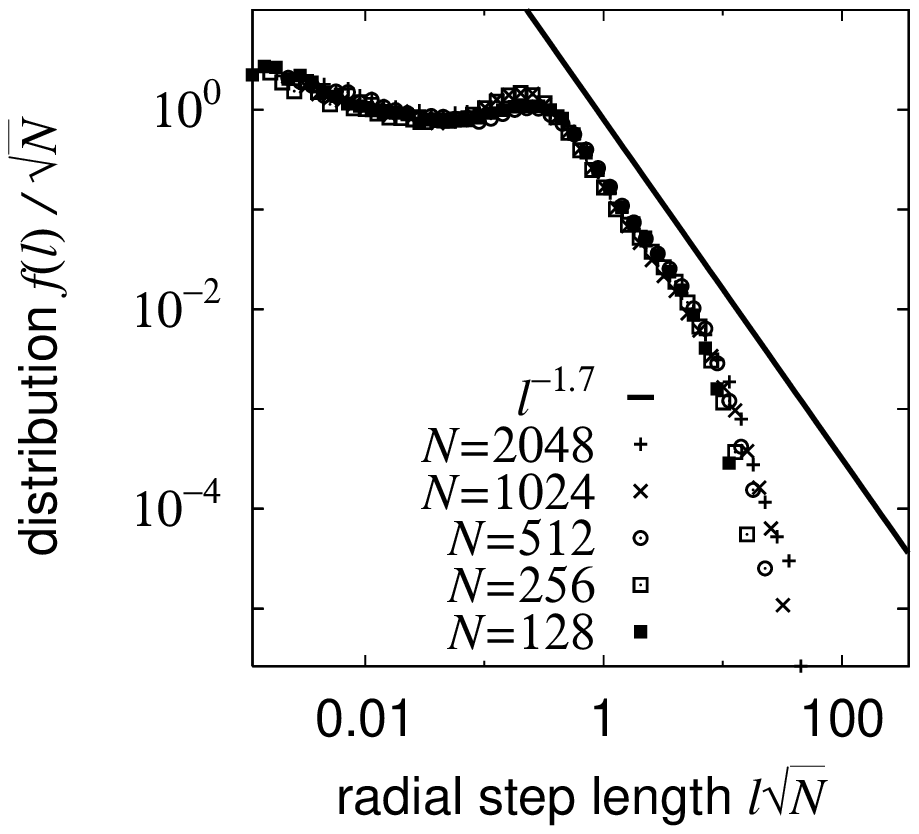}
   \raisebox{0.36\columnwidth}{(b)}
   \hspace{-1em}
   \includegraphics[trim=50 0 50 0,width=0.42\columnwidth]
                   {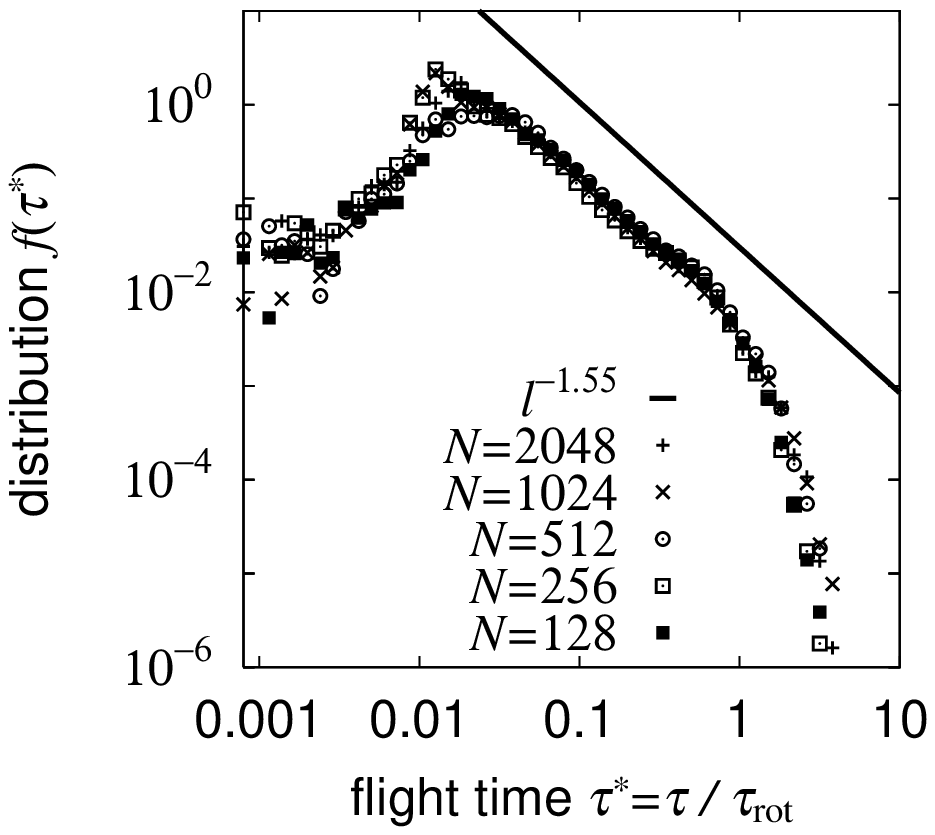}
  \caption{
    Distribution functions for
    the step length $l$ for the radial direction
    scaled by $\sqrt{N}$ (a),  and
    step time $\tau^{*} = \tau / \tau_{\mathrm{rot}}$ in
    units of the bulk rotation time $\tau_{\mathrm{rot}}$ (b).
    The simulation is by Setup A (the pancake state).
    The step length $l$ is defined as a
    distance between two neighboring turns in the trajectories of
    the radial position of a particle,
    and $\tau$ is duration time needed for a step.
    For each simulation with $N$,
    64 particles are randomly chosen and their
    trajectories are analyzed in the slow relaxation,
    $80\tau_{\mathrm{rot}} \leq t \leq 240\tau_{\mathrm{rot}}$.
  }
  \label{f_ps_stepl_logdist}
  \label{f_ps_stepl_logdist-scaled}
  \label{f_ps_ftime_logdist}
 \end{center}
\end{figure}

In order to analyze the above diffusive trajectories as
 shown in Fig. \ref{f_randomwalk},
we decompose them 
into ``steps''. Each step is defined as the
interval between two successive extrema in the radial motion
$\Delta r(t)$,
thus each step has a step length $l$ and a step time $\tau$.
Then the radial position $\Delta r$ and the elapsed time $t$ 
at $M$-th step are expressed by
\begin{equation}
 \label{m_randomwalk_def}
 \Delta r = \sum_{k = 1}^{M}l_{k},
\quad
  t = \sum_{k = 1}^{M}\tau_{k}.
\end{equation}

Figure  \ref{f_ps_stepl_logdist}(a) shows the
distributions of the step length in the simulations of
Setup A (the pancake state) with various $N$,
which exhibits the power law decay
in the large step length.
The step length distributions for various $N$ 
seem to overlap with each other
if we scale the step length as $l\sqrt{N}$, which indicates
that the lower cutoff of the power law scales as 
$l_{\min} \sim 1/\sqrt{N}$.
The upper cutoff of the power law decay is given 
by the system
size,
thus it does not depend on $N$.
The power law distribution is expressed in the form of
$p(l) \sim l^{-(1+\mu)}$ and the 
exponent $\mu = 0.7 \pm 0.1$ is obtained.
We find that the distribution function is almost symmetric both for
the positive $l$ (jump towards the periphery)
 and the negative  $l$(jump towards the center).

Figure \ref{f_ps_ftime_logdist}(b) shows the
similar distribution function obtained for the step time $\tau$,
which 
is normalized by the bulk rotation time
 $\tau_{\mathrm{rot}} \propto 1/N$.
The power law exponent $\chi$ is defined by
$p(\tau) \sim \tau^{-(1 + \chi)}$ and
$\chi = 0.55 \pm 0.1$ is obtained.
The lower cutoff of the power law scales as $\tau_{\min} \sim 1/N$.

\begin{figure}[b]
 \begin{center}
   \raisebox{0.36\columnwidth}{(a)}
   \hspace{-1em}
   \includegraphics[trim=50 0 50 0,width=0.42\columnwidth]
                   {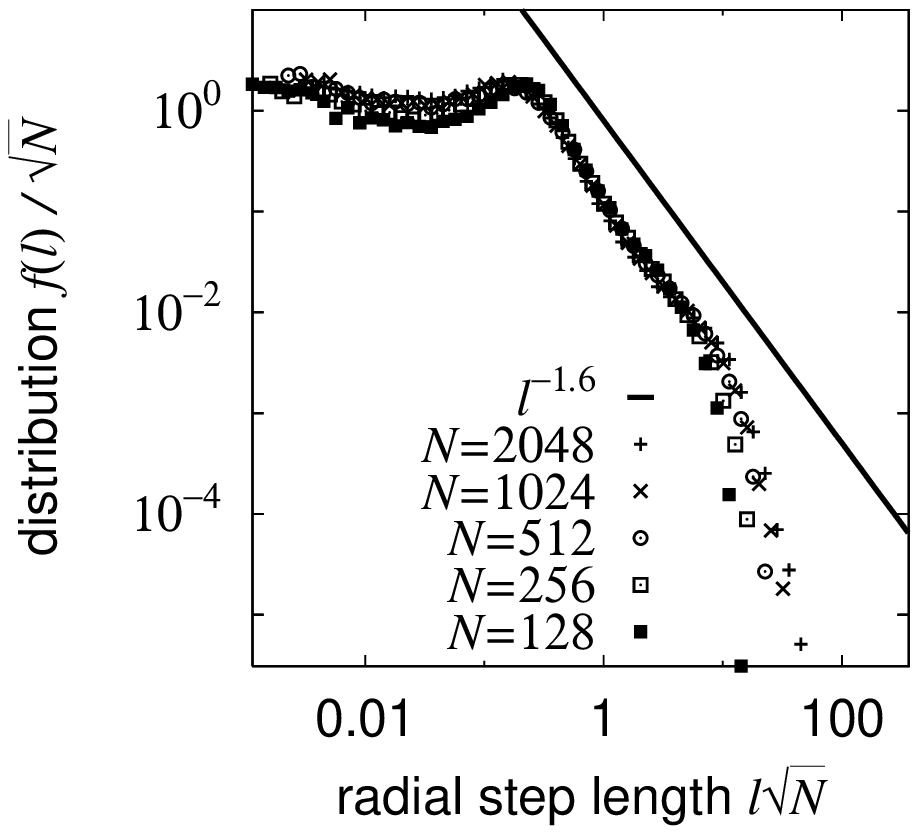}
   \raisebox{0.36\columnwidth}{(b)}
   \hspace{-1em}
   \includegraphics[trim=50 0 50 0,width=0.42\columnwidth]
                   {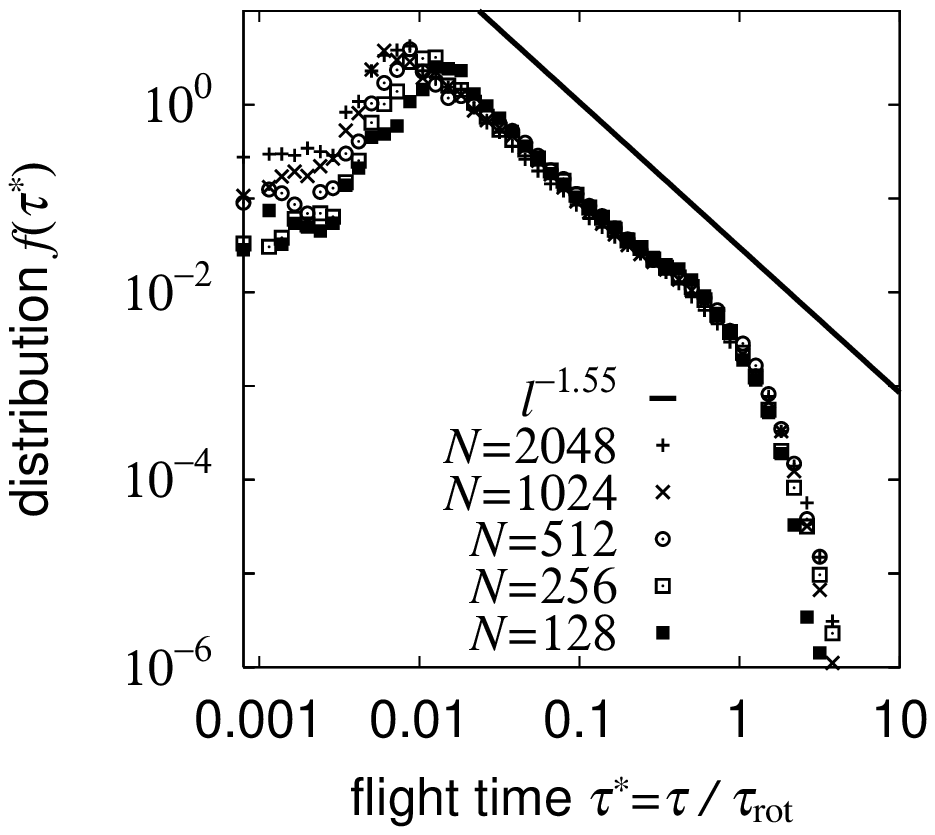}
  \caption{
    Distribution functions for 
    the step length $l$ for
    the radial direction scaled by $\sqrt{N}$  (a),
    and
    the step time $\tau^{*} = \tau / \tau_{\mathrm{rot}}$ in
    the unit of the bulk rotation time $\tau_{\mathrm{rot}}$ (b).
    The simulation is by Setup B (the singly peaked state).
    The data are collected in the same way as in
    Fig. \ref{f_ps_ftime_logdist}
  }
  \label{f_ss_stepl_logdist}
  \label{f_ss_stepl_logdist-scaled}
  \label{f_ss_ftime_logdist}
 \end{center}
\end{figure}

Figure \ref{f_ss_stepl_logdist}
shows the step length distribution $p(l)$ 
and the step time distribution
 $p(\tau)$
for the singly peaked state (Setup B).

Table \ref{texponents} summarizes the exponents
obtained in our simulations;
$\mu$'s and $\chi$'s are the power law exponents of the
step size distributions $p(l)$ and $p(\tau)$, respectively,
and their values are obtained from
 Figs. \ref{f_ps_stepl_logdist} and \ref{f_ss_stepl_logdist}.
The exponents of the anomalous diffusion $\gamma$
from the simulation results
in Fig. \ref{f_ps_anomalous_avgdev},
and
the exponents  $\eta$
for the $N$ dependence of the anomalous diffusion
coefficient $D_{N} \sim N^{\eta}$
from Fig. \ref{f_ps_coeff_anomalous_diff} are also listed.
The exponents $\gamma_{\mathrm{est}}$
and  $\eta_{\mathrm{est}}$
are
the exponents derived from the step distribution exponents
(see \S\ref{s_simulations_ss_analyses}).


\begin{table}[b]
 \begin{center}
  \caption{
   Exponents for
   the step time distributions $\chi$,
   the step size distributions $\mu$,
   and the 
   anomalous diffusion exponents $\gamma$ and $\eta$
   and the derived exponents
   $\gamma_{\mathrm{est}}=2\chi/\mu$
   and 
   $\eta_{\mathrm{est}}=2\chi/\mu-1$
   from $\chi$ and $\mu$
   See texts for the definition of each exponent.
  }
  \label{texponents}
  \begin{tabular}{|c|c|c|c|c|}
   Setup
    & $\chi$ & $\mu$
    & $\gamma$ & $\eta$ \\
   \hline
   A 
    & $0.55 \pm 0.1$ & $0.7 \pm 0.1$
    & $1.75 \pm 0.1$ & $0.85 \pm 0.1$ \\
   B 
    & $0.55 \pm 0.1$ & $0.6 \pm 0.1$
    & $1.85 \pm 0.1 $ & $1.2 \pm 0.1$ \\
   \hline
   Setup
    & &
    & $\gamma_{\mathrm{est}}$
    & $\eta_{\mathrm{est}}$ \\
   \hline
   A 
    & &
    & $1.6$ & $0.6$ \\
   B 
    & &
    & $1.8$ & $0.8$ \\    
  \end{tabular}
 \end{center}
\end{table}


\begin{figure}[b]
 \begin{center}
   \raisebox{0.36\columnwidth}{(a)}
   \hspace{-1em}
   \includegraphics[trim=50 0 50 0,width=0.42\columnwidth]
                   {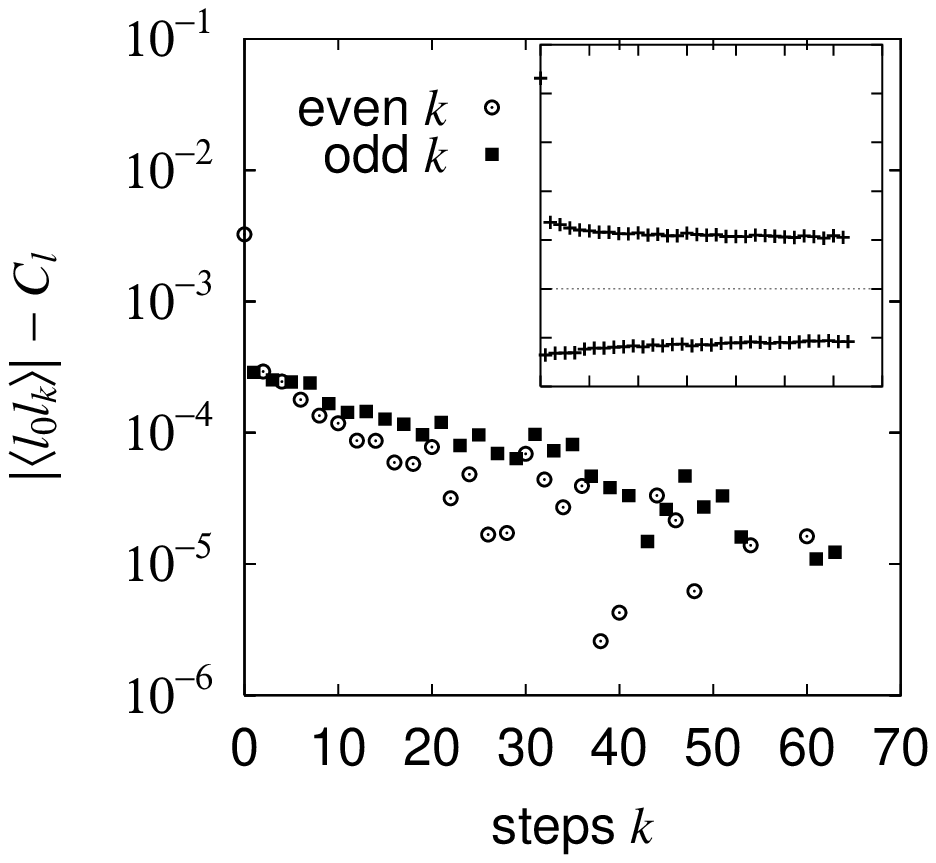}
   \raisebox{0.36\columnwidth}{(b)}
   \hspace{-1em}
   \includegraphics[trim=50 0 50 0,width=0.42\columnwidth]
                   {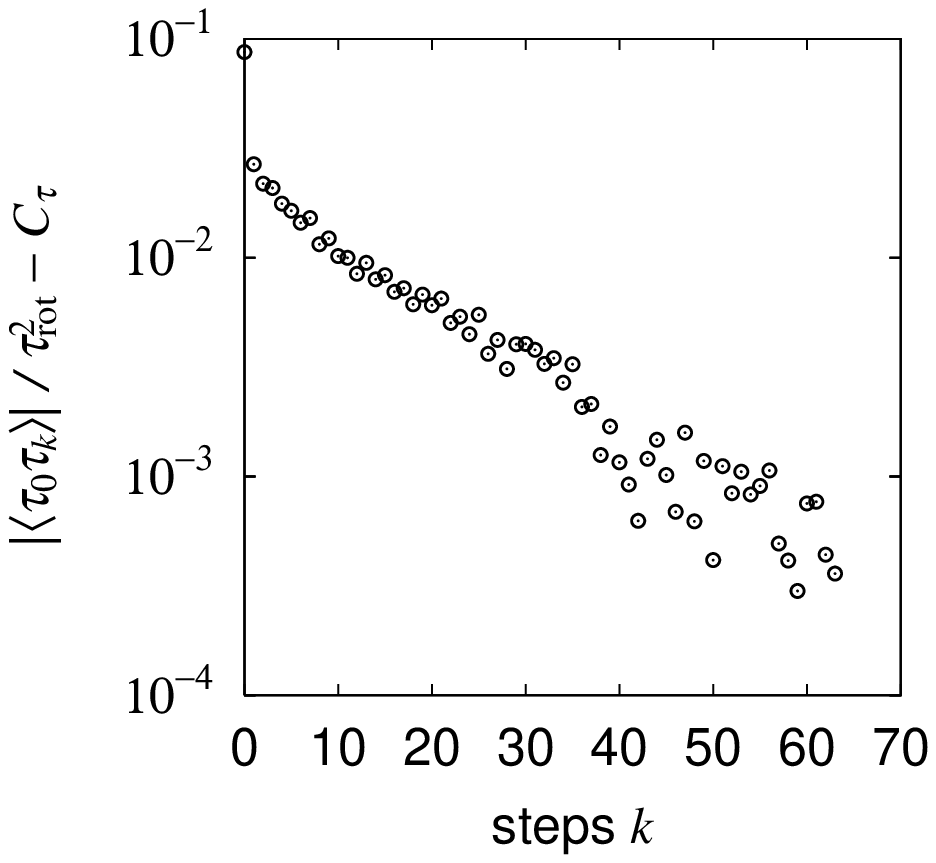}
  \caption{
    Correlation functions in the step sequences of the
    step length $\langle l_{0}l_{k}\rangle$ (a), and
    the step time $\langle \tau_{0}\tau_{k}\rangle$ (b)
    as a function of $k$.
    The simulation is by Setup A (the pancake state)
    with $N = 2048$.
    These data are obtained from trajectories of randomly chosen
    64 particles in the slow relaxation process,
    $80\tau_{\mathrm{rot}} \leq t \leq 240\tau_{\mathrm{rot}}$.
    The fitting parameters are 
    $C_{l} = 0.00107$ and $C_{\tau} = 0.539$.
    The inset in (a) shows a linear plot of step correlation
    in the same range of $k$.
  }
  \label{f_ps_steplcorrelation}
 \end{center}
\end{figure}

Now we examine the correlation in the step sequences.
Let the step length and step time of the $k$-th step be
 $l_{k}$ and $\tau_{k}$, respectively.
The correlations
$\langle l_{0}l_{k}\rangle$ and
$\langle \tau_{0}\tau_{k}\rangle$
are plotted against
 $k$ for the pancake state
(Setup A)
in Fig. \ref{f_ps_steplcorrelation}.
These correlations show the exponential decay,
which indicates that the correlation in the step sequences
is short-range.

\begin{figure}[b]
 \begin{center}
   \raisebox{0.34\columnwidth}{(a)}
   \hspace{-1em}
   \includegraphics[trim=60 0 35 0,width=0.42\columnwidth]
                   {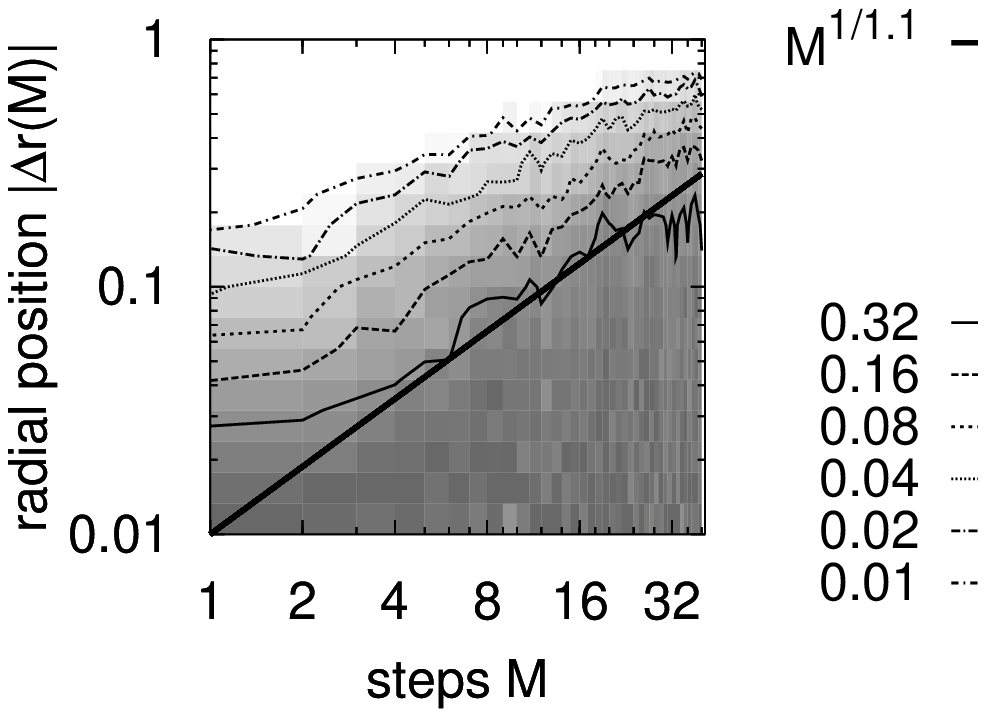}
   \raisebox{0.34\columnwidth}{(b)}
   \hspace{-1em}
   \includegraphics[trim=60 0 35 0,width=0.42\columnwidth]
                   {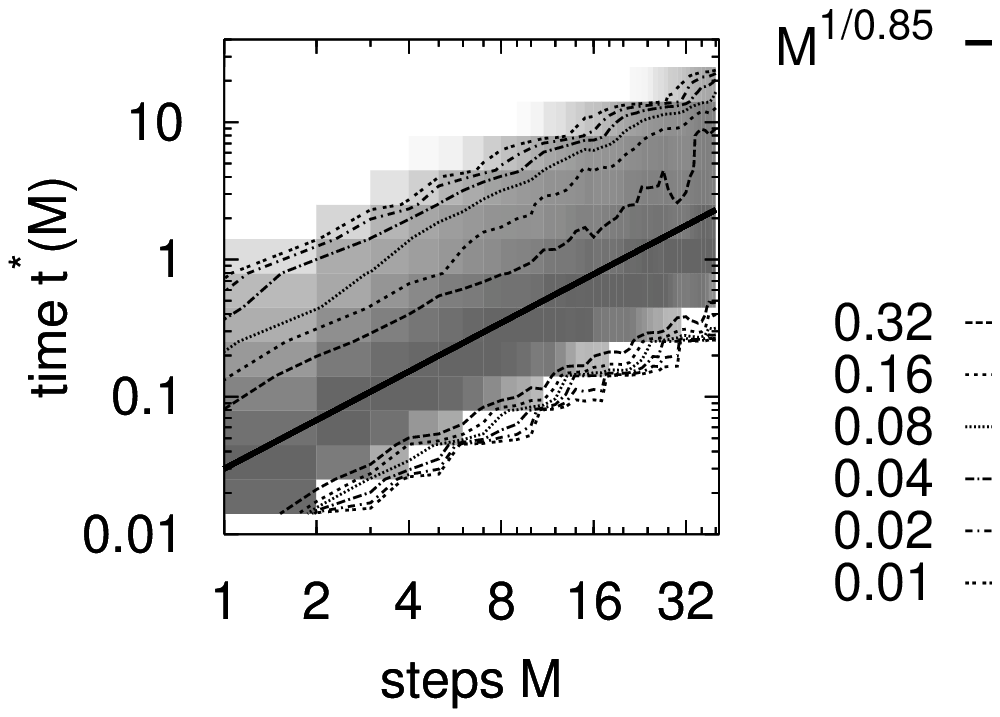}
  \caption{
   Contour plots for the distribution of
   radial position  $P(|\Delta r|\, \bigl| M)$ (a), and
   that of the time $P(t^{*}|M)$ (b)
   as a function of the number of steps $M$ in the logarithmic scale.
   $t^{*} = t/\tau_{\mathrm{rot}}$ and
   the distribution functions are normalized by the peak distribution
   $p_{\mathrm{peak}}(M)$ for each step $M$.
   The thick lines are $|\Delta r| \propto M^{1/\mu^{*}}$ (a)
   and $t^{*} \propto M^{1/\chi^{*}}$ (b)
   with  $\mu^{*} = 1.1$ and  $\chi^{*} = 0.85$ that are obtained
   by fitting.
    The data are obtained from trajectories of randomly chosen
    64 particles in the time duration,
    $80\tau_{\mathrm{rot}} \leq t \leq 240\tau_{\mathrm{rot}}$.
   For both figures, the
   simulation is by Setup A (the pancake state) with $N=2048$.
  }
  \label{f_ps_logmap-r}
 \end{center}
\end{figure}

In order to analyze the step sequence
in more detail, we define $P(|\Delta r|\, \bigl| M)$
and $P(t|M)$ as the distributions of
the distance $\Delta r$ and the time $t$
at the  $M$-th step (eq. (\ref{m_randomwalk_def})),
respectively.
Contour plots of these distribution functions
are shown in Fig. \ref{f_ps_logmap-r}.
The value of the distributions are shown
 in the proportion to the peak value
$p_{\mathrm{peak}}(M)$ of $P(|\Delta r|\, \bigl| M)$ or $P(t | M)$
for each step $M$
so that one can compare the width of the distributions between
different $M$'s.
We observe that the width of $P(|\Delta r|\, \bigl| M)$ and the
peak position of $P(t | M)$
increase by the powers of $M$, as
\begin{equation}
\label{m_map_anomalous_diff}
(\Delta r)^{2} \sim M^{2/\mu^{*}},
\quad
t \sim M^{1/\chi^{*}}
\end{equation}
with $\mu^{*} = 1.1 \pm 0.2$ and
$\chi^{*} = 0.85 \pm 0.1$.

Note that the contour curves are almost equally spaced
in the logarithmic scale,
which means that
 the distribution functions have the power-law tail.
The growth of the
tail of the distribution function $P(|\Delta r|\, \bigl| M)$ 
is saturated for large $M$
 where the tail approaches $\Delta r \sim 1$, or 
the system size.

\section{Analyses}
\label{s_simulations_ss_analyses}
As we have shown in \S\ref{s_simulations_ss_results_sss_randomwalk},
the step length distribution $p(l)$ and the step time
distribution $p(\tau)$ have the power law tail,
and the correlation between the steps
decays exponentially.
Based on these observations, we analyze the diffusion process
in terms of the Levy flight and derive the exponents.
Some of the basic properties of the Levy flight are summarized in
 Appendix.

The typical radial deviation of position $\Delta r_{M}$ and
the typical elapsed time $t_{M}$
after the $M$-th step can be estimated by Levi flight as
\begin{equation}
\label{m_position_anomalous_diff}
 \Delta r_{M}^{2}
 \sim
 M^{2/\mu}\, l_{\min}^{2},
\quad
t_{M}
 \sim 
  M^{1/\chi}\, \tau_{\min},
\end{equation}
from  eq. (\ref{mlevy_typicalwalk}).

These should be compared with 
eq. (\ref{m_map_anomalous_diff}), which are
direct observation of $P(\Delta r | M)$ and $P(t|M)$.
Both of them show the power law of $M$, but the
values of the exponents $\mu^{*}$ and $\chi^{*}$
estimated from the direct observation are not very close
to the values of $\mu$ and $\chi$ from the
step distribution. we suspect this discrepancy is
due to the finite size effect as we have mentioned at
the end of \S\ref{s_simulations_ss_results_sss_randomwalk}.

Keeping this limitation in mind,
let us continue the analysis based on the Levy flight picture;
The expression (\ref{m_position_anomalous_diff})
show
that the total sums of eq. (\ref{m_randomwalk_def})
are of the same order of the largest step out of the $M$ steps.
This means that most of the contribution to
 eq. (\ref{m_randomwalk_def}) comes from the largest step.
We assume this largest step to be the same step for 
$\Delta r$ and $t$. This is natural assumption
because shorter steps are results of 
waggling motion caused by nearly vortices,
vortices tend to move faster during the shorter steps.


Using the assumption,
the typical radial position $\Delta r_{t}$ at time $t$ can
be estimated by simply eliminating $M$ in
eq. (\ref{m_position_anomalous_diff}).
Identifying $t_{M}$ as $t$,
we have
\begin{equation}
\label{m_levy_rt}
\Delta r_{t}^{2}
 \sim
  D_{N}t^{2\chi/\mu}
\end{equation}
with
\begin{equation}
\label{manomalous_diffusion_coefficient_estimate}
D_{N} \sim
  \frac{l_{\min}^{2}}{\tau_{\min}^{2\chi/\mu}} .
\end{equation}
Comparing eq. (\ref{m_levy_rt}) with
eq. (\ref{m_anomalous_diff}),
 we see that the anomalous diffusion
exponent $\gamma$ is related with $\chi$ and $\mu$ as
\begin{equation}
\label{m_estimate_gamma}
\gamma = 2\chi / \mu.
\end{equation}

The dependence of the anomalous diffusion coefficient $D_{N}$
on the number of particles $N$ is also obtained from
eq. (\ref{manomalous_diffusion_coefficient_estimate}).
Assuming that the short length
 cutoff of the step length distribution
$l_{\min}$ is approximately given by the average particle distance $l_{0}$,
thus we have
\begin{equation}
l_{\min}^{2} \approx l_{0}^{2} \sim n_{0}^{-1} = \frac{\pi L^{2}}{N}
\end{equation}
Similarly, we assume that
the short time cutoff of the step time distribution $\tau_{\min}$
is given by the rotation time of the vortex pair $\tau_{0}$
with the distance $l_{0}$:
\begin{equation}
\tau_{\min} \approx \tau_{0} \sim \frac{l_{0}}{v_{0}} 
 \sim N^{-1}
\end{equation}
where $v_{0}$ is the rotation velocity of the paired vortices 
with the distance $l_{0}$, and given by
\begin{equation}
v_{0} \sim  \frac{q}{l_{0}},
\end{equation}
from the 2-d Coulomb law. Here, $q$ is the
charge of a particle, which is normalized to unity.
Note that these $N$ dependences are consistent with 
the scaling form of the step size distributions $p(l)$
and $p(\tau)$ in
Figs. \ref{f_ps_stepl_logdist} and \ref{f_ss_stepl_logdist}.
Substituting these cutoffs into
 eq. (\ref{manomalous_diffusion_coefficient_estimate}),
we obtain $N$ dependence of the coefficient $D_{N}$,
\begin{equation}
\label{m_estimate_eta}
 D_{N} \sim N^{2\chi/\mu - 1},
\end{equation}
thus the exponent $\eta$ as
\begin{equation}
\eta = 2\chi / \mu - 1 = \gamma - 1 .
\end{equation}

From the simulation results and the definition of the exponents,
we obtain $\chi = 0.55$ and $\mu = 0.7$ for the
radial direction in the pancake state (Setup A).
Thus, from eqs. (\ref{m_estimate_gamma}) and (\ref{m_estimate_eta}),
one obtains
 $\gamma \approx 1.6$ and $\eta \approx 0.6$. 
Similarly, we obtain the exponents for the singly peaked
state (Setup B). They are listed in Table \ref{texponents}.

\section{Summary and discussion}
\label{s_conclusions_and_discussion} 
We have performed the numerical simulations on the two-dimensional point
vortex model with a unit circulation of the same sign,
 in order to study the relaxation process of
pure electron plasma under the strong magnetic field.
Due to the long-range interaction between the vortices,
the system behaves very differently from ordinary systems
with short-range interaction.

We have found the following:
(i) 
There exist two types of relaxation: the fast relaxation and the slow
relaxation. 
The fast relaxation takes place with the time scale comparable with
the bulk rotation time $\tau_{\mathrm{rot}}$
and leads the system to a 
quasi-stationary state following the Euler equation,
 while
the slow relaxation takes place 
after the fast relaxation with the relaxation time
$\tau_{\mathrm{relax}} \sim N\tau_{\mathrm{rot}}$
due to the individual motion of vortices;
This is
 consistent with Chavanis' theory except for the
$\log N$ correction.
After the slow relaxation,
the system reaches the maximum one-body entropy state.
(ii)
Individual motion of point vortices in
the slow relaxation process is
superdiffusive with the exponent $\gamma \approx 1.75$
for the pancake (shear free) state and $\gamma \approx 1.85$
for the singly peaked (with shear) state.
The coefficient of anomalous diffusion depends on the number
of particles $N$ in the power laws.
(iii) 
The superdiffusive motion of individual
vortices can be decomposed into a 
sequence of steps.
The correlations of the step length and that of the step time
along the sequence are short range, and the 
distributions of the step length and the step time are
of the power laws with the exponents $\mu$ and $\chi$, respectively.
The superdiffusive motion can be reconstructed from the
Levy flight, i.e.
the exponent $\gamma$ for the anomalous diffusion
and the exponent $\eta$ for the $N$ dependence of 
its coefficient
are expressed as $\gamma = 2\chi/\mu$ and 
$\eta = 2\chi / \mu - 1$;
the former expression agrees with the exponent from the
simulation within the error bars,
 but the agreement of the latter is not good.

%

Among these results, 
our result (i) of
the $N$ dependence of the slow relaxation time
seems to agree with Chavanis' estimation. However, his picture
that the relaxation is due to the normal diffusion 
of the point vortices is not confirmed by our simulation results,
but
we observe the superdiffusive behavior in the simulations.
If we estimate the relaxation time $ \tau_{\mathrm{relax}}^{*}$
with anomalous diffusion
by the similar way as he did for normal diffusion, then
\begin{equation}
 \tau_{\mathrm{relax}}^{*}
\sim
 \left(\frac{L^{2}}{D_{N}}\right)^{1/\gamma}
\sim 
 N^{1 - \eta/\gamma}\tau_{\mathrm{rot}}
\ll 
 \tau_{\mathrm{relax}} \sim N \tau_{\mathrm{rot}},
\end{equation}
which gives much shorter relaxation time than that observed.
This implies that the motions of point vortices 
are not independent of each other and
provide only weak mixing.
Actually we observe in our simulations that
a point vortex tends to avoid ``collisions''
during the long jumps.


In the literature,
\cite{
 bplasma_diffusion_two_dimensions,
 bnumerical_simulation_plasma_difusion_across
}
it has been pointed out that
a test particle is convected for a long way comparable to
the system size due to the existence of long-living large vortices.
The diffusion coefficient proposed by
Taylor and McNamara depends on time and
converges to a constant
only when $(\Delta r / L)^{2} \gg 1$ holds,
which suggests anomalous diffusion.
Although these results are for 
the neutral plasma,
its behavior of the anomalous diffusion
seems consistent qualitatively with our simulation results
on the non-neutral plasma.

Recently,
Dubin and Jin performed fairly large-scale
simulations on the 2-d point vortex system with a positive charge,
\cite{
 bcollisional_diffusion_one,
 bcollisional_diffusion_multi
}
and determined the diffusion coefficients
in the states without mean shear.
Their results of the diffusion coefficient
show the $N^{1/2}$ dependence, which is consistent
with that expected by
Taylor and McNamara in the converging limit,
but their values of diffusion constant from the simulations
seem substantially smaller than those
predicted by the theory:
This may be due to the slow convergence mentioned in the above
paragraph.
They have also shown that the diffusion in the radial direction
is normal under the external shear.

In experiments,
the number of electrons is very large while the charge of each
electron is very small.
In the present model, the situation may correspond
to the case in the limit of the infinite $N$
with a fixed total charge, namely, a fixed bulk
rotation time  $\tau_{\mathrm{rot}}$,
in which limit the slow relaxation never takes place
and we only observe the quasi-stationary states.
If this is the case, actual slow relaxation that may be observed
in experiments should be due to
 non-ideal effects such as three dimensionality
of the Malmberg trap, scattering at the end of the cylinder,
 impurities, etc.

%

\section*{Acknowledgments}
The authors would like to thank 
 Professor Y. Kiwamoto, Professor M. Sakagami,
Dr. M. M. Sano, Dr. Y. Yatsuyanagi, T. Yoshida and Dr. D. Watanabe for
 valuable discussions and comments.
They would also like to thank Professor P. H. Chavanis
for his communication.

\appendix

\section{Levy flight and anomalous diffusion}
\label{s_levi}
In this appendix, we summarize some basic formulas of the
Levy flight.

Consider a random walk problem where the size of each step $l$
is a random variable without correlation,
 then after the $M$-th step the particle position $X$ is
\begin{equation}
X = \sum_{i}^{M} l_{i} .
\end{equation}
If the distribution $p(l)$ of step size $l$ has a finite
second moment $\langle l^{2} \rangle$,
the central limit theorem tells us that
the distribution of $X$ at the $M$-th step
is known to be the Gaussian distribution for large $M$
and the dispersion increases as 
\begin{equation}
\langle (X - \langle X \rangle)^{2} \rangle
\sim D M,
\end{equation}
where $D$ is the diffusion coefficient.
This represents the normal diffusion law.

If the distribution function $p(l)$ is the power law
with the diverging second moment,
\begin{equation}
p(l) \sim |l|^{-(1+\mu)},
\quad
 (0 < \mu < 2) 
\end{equation}
for large $|l|$,
then the asymptotic form of the distribution of $X$
at large $M$ is given  by the
Levy's stable distribution.
\cite{
 banomalous_diffusion_disordered,
 blimit_distributions_sums}
\begin{equation}
P(X|M) \rightarrow
\left\{
\begin{array}{ll}
 \frac{1}{M^{1/\mu}}
  L_{\mu, \beta}\left(\frac{X}{M^{1/\mu}}\right),
 &
 (0 < \mu < 1), \\
 \frac{1}{M^{1/\mu}}
  L_{\mu, \beta}\left(\frac{X - \langle X \rangle}{M^{1/\mu}}\right),
 &
 (1 < \mu < 2),
\end{array}
\right.
\end{equation}
where $L_{\mu, \beta}(x)$ is the scaling form of the distribution of
 $X$ at $M \rightarrow \infty$.
The parameter $\beta$ characterizes the degree of asymmetry,
and is determined from the asymmetry of $p(l)$ for large $|l|$.
The $\beta = 0$ case represents the symmetric distribution:
\begin{equation}
 L_{\mu, 0}(Z)
  \equiv
 \frac{1}{2\pi}\int_{-\infty}^{+\infty}dk \exp(i kZ-C|k|^{\mu}) ,
\end{equation}
where $C$ is a scaling factor.
This 
gives the Gaussian distribution when $\mu = 2$.
As for the $\beta = 1$ case, the distribution is given by
\begin{equation}
 L_{\mu, 1}(Z)
  \equiv
 \frac{1}{2\pi i}\int_{d -i\infty}^{d+i\infty}ds \exp(sZ-C's^{\mu}),
\end{equation}
which is zero for $Z < 0$ representing a completely asymmetric case.
The distribution  $L_{\mu, \beta}(Z)$ 
has an approximate form of
$L_{\mu, \beta}(Z) \sim Z^{-(1+\mu)}$ for large $Z$.

The fact that the above distribution function $P(X|M)$
scales as $X/M^{1/\mu}$
can be understood\cite{banomalous_diffusion_disordered}
by introducing the effective cutoff
 $l_{\mathrm{c}}$,
to the step length distribution $p(l)$.
By simple argument, one can see the cutoff depends on the number of
steps $M$ as
\begin{equation}
l_{\mathrm{c}} \sim M^{1/\mu}l_{\min} ,
\quad
(0 < \mu < 1) .
\end{equation}
Then, a typical walk may be estimated as
\begin{equation}
\label{mlevy_typicalwalk}
X_{M} = \sum_{i}^{M}l_{i} \approx M\langle l \rangle
\approx \int_{l_{\min}}^{l_{\mathrm{c}}} dl\, l\, p(l)
\sim M^{1/\mu}l_{\min},
\quad
(0 < \mu < 1) ,
\end{equation}
which means the distribution function scales as $X/M^{1/\mu}$.
Therefore, 
peak position $X_{\mathrm{p}} \sim M^{1/\mu}$
 and the square of typical width
 $X_{\mathrm{w}}^{2} \sim M^{2/\mu}$
 of the distribution $P(X|M)$
at the $M$-th step can be considered as measures of the anomalous diffusion
by the Levy flight,
 even though the distribution function has
diverging moments.

%

\bibliographystyle{jpsj}
\bibliography{slowrelax_ronbun}

\end{document}